\documentclass[
    aps,
    prb,
    physrev,
    twocolumn,
    groupedaddress,
    superscriptaddress,
    footinbib,
    floatfix,
    longbibliography,
    showpacs,
    ]{revtex4-2} 

\usepackage[utf8]{inputenc}
\usepackage{amssymb,amsmath,mathtools,graphicx,xcolor}
\usepackage{bm} %

\usepackage[bookmarksnumbered,              %
colorlinks = true,
linkcolor = black,
urlcolor  = blue,
citecolor = blue,
anchorcolor = blue]{hyperref}

\usepackage{multirow} %
\usepackage{booktabs} %
\usepackage{import} %
\usepackage{xstring} %
\usepackage{bbm} %
\usepackage[capitalise]{cleveref} %
\crefname{appendix}{App.}{Apps.}

\usepackage{orcidlink} %
\graphicspath{{figures}}

\DeclarePairedDelimiterX\abs[1]{\lvert}{\rvert}{#1}
\DeclarePairedDelimiterX\ket[1]{\lvert}{\rangle}{#1}
\DeclarePairedDelimiterX\bra[1]{\langle}{\rvert}{#1}
\DeclarePairedDelimiterX\braket[2]{\langle}{\rangle}{#1\vert#2}
\DeclarePairedDelimiterX\ketbra[2]{\lvert}{\rvert}{#1\rangle\langle#2}
\DeclarePairedDelimiterX\projector[1]{\lvert}{\rvert}{#1\rangle\langle#1}
\DeclarePairedDelimiterX\expval[2]{\langle}{\rangle}{#1\vert#2\vert#1}
\DeclarePairedDelimiterX\matel[3]{\langle}{\rangle}{#1\,\vert\,#2\,\vert\,#3}
\DeclarePairedDelimiterX\ep[1]{\langle}{\rangle}{#1}
\DeclarePairedDelimiterX\im[1]{\mathrm{Im}\lbrack }{\rbrack}{#1}
\DeclarePairedDelimiterX\re[1]{\mathrm{Re}\lbrack }{\rbrack}{#1}

\newcommand{\del}{\partial}
\newcommand{\dg}{\dagger}
\newcommand{\vac}{\ket{\text{vac}}}
\newcommand{\zv}{\bm{0}}
\newcommand{\dv}{\bm{d}}
\newcommand{\rv}{\bm{r}}
\newcommand{\kv}{\bm{k}}
\newcommand{\colvec}[1]{\begin{pmatrix}#1 \end{pmatrix}}
\newcommand{\rowvec}[1]{\begin{pmatrix}#1 \end{pmatrix}}
\newcommand{\diag}{\text{diag}\colvec}

\newcommand{\ccite}[1]{%
\IfSubStr{#1}{,}{refs.}{ref.~}\cite{#1}%
}
\newcommand{\Ccite}[1]{%
\IfSubStr{#1}{,}{Refs.~}{Ref.~}\cite{#1}%
}

\begin{document}

\newcommand{\titlecontent}{
Symplectic Hopf Insulator: \\ Delicate Topology in Bosonic Bogoliubov-de Gennes Systems}
\title{
    \titlecontent
    }

\newcommand{\ITPTUB}{Institut f\"{u}r Physik und Astronomie, Technische Universit\"{a}t Berlin,
Hardenbergstr.~36, D-10623 Berlin, Germany}

\author{Isaac Tesfaye\,\orcidlink{0009-0001-4194-3916}}
\email{i.tesfaye@tu-berlin.de}
\affiliation{\ITPTUB}

\author{Giandomenico Palumbo\,\orcidlink{0000-0003-1303-1247}}
\email{giandomenico.palumbo@gmail.com}
\affiliation{CFisUC, Department of Physics, University of Coimbra, Rua Larga, 3004-516 Coimbra, Portugal}

\begin{abstract}
Recent advances in topological phases have highlighted the role of symplectic (Krein-space) topology in the classification of bosonic Bogoliubov-de Gennes (BBdG) systems. In this work, we construct a BBdG realization of Hopf topology, which we dub the symplectic Hopf insulator, starting from a microscopic Bose-Hubbard generalization of the Moore-Ran-Wen model with weak on-site interactions treated within a Bogoliubov approximation. The resulting BBdG system admits a symplectic Hopf invariant, which we show to be integer-quantized for isolated bands. 
We establish that this topology is intrinsically delicate, requiring exactly two bosonic modes per unit cell, while remaining robust against weak interactions over a range of mass parameters.
Upon terminating the three-dimensional insulator at a boundary, we find topologically protected in-gap surface states at finite excitation energy, whose protection is itself delicate. Our results establish the symplectic Hopf insulator as a robust yet delicate topological phase in weakly interacting bosonic systems lying beyond the tenfold-way classification.
\end{abstract}

\date{\today} %

\maketitle

\makeatletter
\let\origaddcontentsline\addcontentsline
\renewcommand{\addcontentsline}[3]{}
\makeatother

\hypersetup{linkcolor=blue} %

\section{Introduction}
\label{sec:introduction}
Topological band insulators and superconductors~\cite{Qi2011,Hasan2010} are a cornerstone of modern condensed matter physics, which can be classified by the so-called tenfold-way~\cite{Zirnbauer1996,Altland1997,Schnyder2008,Kitaev2009a,Ryu2010} (including crystalline symmetries if present~\cite{Chiu2016}) based on stable equivalence in $K$-theory~\cite{Karoubi1978,Kitaev2009a,Freed2013}. 
There are, however, topological phases that lie outside this classification, one of which is the paradigmatic Hopf insulator~\cite{Moore2008}, which is a three-dimensional topological phase governed by a topological invariant, called the Hopf index~\cite{Hopf1931,Whitehead1947,Wilczek1983,Pontryagin1941,Ren2007,Moore2008}. 
Unlike the stable tenfold-way topological phases, which allow for the addition of trivial bands to the conduction and valence bands above and below the band gap, respectively,
and also unlike fragile topology~\cite{Po2018,Bradlyn2017,Bouhon2020}, which is nullified only by adding trivial bands to the valence subspace, the Hopf insulator is a so-called \emph{delicate} topological phase, which necessitates exactly two bands and is unstable against the addition of trivial bands to either the conduction or valence band~\cite{Nelson2021,Lapierre2021,Nelson2022,Brouwer2023}.
Mathematically, it is classified by the homotopy invariant $\pi_3(S^2) = \mathbb{Z}$~\cite{Kennedy2014,Kennedy2015,Kennedy2016} with the two-sphere $S^2$ as the state space of the two-band insulators. 
Various special features of the Hopf insulator, unlike those of its stable counterparts, have been revealed in recent years~\cite{Deng2013,Kennedy2016,Liu2017a,Alexandradinata2021,Zhu2021,Zhu2023a}. The Hopf insulator has recently been realized in a circuit-QED quantum simulator~\cite{Wang2023} and proposed for dipolar spin systems~\cite{Schuster2021a}, making its features experimentally accessible. 
Other theoretical extensions of the Hopf insulator include Floquet Hopf insulators~\cite{Unal2019,Schuster2019,He2019}, connections to the Hopf-Euler models~\cite{Jankowski2024b,Lim2023}, non-Hermitian variants~\cite{Yang2019,He2020,Nakamura2025,Yoshida2026}, an $N$-band generalization~\cite{Lapierre2021},
and a Hopf superconductor model~\cite{Kennedy2016}, which is most related to this work.
In particular, Kennedy~\cite{Kennedy2016} completed the class-$A$ classification of two-band models in three dimensions, where the Hopf invariant is not integer-quantized once the weak sub-tori (first) Chern numbers do not all vanish, giving rise to the so-called Hopf-Chern insulators. 
The fermionic Hopf superconductor generalizes this to a Bogoliubov-de~Gennes (BdG) setting where, because of the fermionic anti-commutation relations, the Bogoliubov transformation is \emph{unitary}, so that its classification stays close to that of free fermions.

On the other hand, bosonic BdG (BBdG) systems, due to the bosonic commutation relations, are diagonalized by so-called paraunitary (symplectic) Bogoliubov matrices~\cite{Colpa1978}.
The topological features of such BBdG systems have been well studied~\cite{Tesfaye2025,Shindou2013,Shindou2013a,Furukawa2015,Wang2021a,Jalali-mola2023,Engelhardt2015,Kondo2019,Lieu2018,Ohashi2020,Bardyn2016,Peano2016a,Peano2016,Peano2018,Lein2019,McDonald2018,Okuma2023a,Wan2021,Chaudhary2021,Xu2020,Massarelli2022,DiLiberto2016a,Flynn2020a,Huang2021a,Huang2022a,Chen2023a,Ravets2025,Julku2021,Julku2021a,Julku2023,Salerno2023,Zhou2020,Gong2018,Kawabata2019,Zhou2019,Bernard2002,Yuan2026,Yamamoto2026}. These systems support bosonic chiral edge magnons and edge matter waves at finite energy.
This paraunitary structure renders the effective single-particle matrix, the so-called dynamical matrix pseudo-Hermitian, so that BBdG bands can even host genuinely non-Hermitian features such as exceptional points and dynamical instabilities that are absent in the fermionic case~\cite{Zhang2019,Zhu2021,ChenYe2024}. 
The three-dimensional BBdG topological classification beyond the tenfold way~\cite{Zhou2020,Gong2018,Kawabata2019,Zhou2019,Lieu2018,Bernard2002}, and, in particular, a BBdG counterpart of the delicate Hopf insulator, is, however, entirely unexplored.
This raises the question we want to address here:
What happens to the Hopf topology when the underlying model admits a bosonic BdG structure, with a paraunitary Bogoliubov transformation acting on an indefinite Krein inner product space spanned by the Bogoliubov modes?

To answer this, we construct a symplectic Hopf insulator model by starting from a two-sublattice Bose-Hubbard Moore-Ran-Wen model generalization on a cubic lattice and applying the Bogoliubov approximation for weak interactions to obtain a BBdG Hamiltonian. 
Using the paraunitary structure of the Bogoliubov transformation, we define the \emph{symplectic Hopf invariant}, which we show to be integer-quantized for isolated BBdG bands.

The remainder of the paper is organized as follows.
In \Cref{sec:hermitian-hopf} we review the two-band Moore-Ran-Wen (MRW) Hopf-insulator model on
which our construction builds.
\Cref{sec:symplectic-hopf-model} introduces the symplectic (BBdG) Hopf insulator model, where in
\Cref{subsec:symplectic-hopf-invariant} we construct the symplectic Hopf invariant. 
In~\Cref{subsec:Phase-Diagram} and~\Cref{subsec:edge-states} we present the phase diagram for the symplectic Hopf invariant at finite interaction strengths and the emergence of the delicate in-gap surface states upon opening the system on a boundary, respectively. 
We summarize our findings and discuss open questions in \Cref{sec:summary-outlook}.

\section{Recap: Hermitian Hopf insulator}
\label{sec:hermitian-hopf}

We review the seminal three-dimensional (3D) two-band Moore-Ran-Wen (MRW) Hopf insulator model~\cite{Moore2008}. 
The two-band Bloch Hamiltonian of the MRW model is given by
\begin{align}
    \mathcal{H}(\kv) = \bm{d}(\kv)\cdot\bm{\sigma},
    \quad
    d_i(\kv) = \expval{u}{\sigma_i}, \ i\in\{1,2,3\},
    \label{eq:mrw-hamiltonian}
\end{align}
where $\bm{\sigma}=(\sigma_1,\sigma_2,\sigma_3)$ is the vector of Pauli matrices, $\dv(\kv) = (d_1(\kv),d_2(\kv),d_3(\kv))$ the Bloch vector, with eigenvalues $E_\pm(\kv) = \pm\abs{\bm{d}(\kv)}$. 
It is built from the two-component spinor $\ket{u}=(u_1,u_2)^T$ with $u_a \in \mathbb{C}, a \in \{1,2\}$ given by
    \begin{align}
        u_1(\kv) &= \sin k_x + i\sin k_y, \nonumber \\
        u_2(\kv) &= \sin k_z + i (\cos k_x + \cos k_y + \cos k_z + m),
    \end{align}
where $\kv=(k_x,k_y,k_z)\in T^3$ is the crystal momentum, and $m$ the mass parameter tuning the topological phase transition of the conventional Hopf index below.
Normalizing the spinors via  $\ket{z} = \ket{u}/\sqrt{\braket{u}{u}}$ together with 
$R_i = \expval{z}{\sigma_i} = \hat{d}_i = d_i/\abs{\bm{d}}$ gives us the composite map $T^3 \xrightarrow{z} S^3 \xrightarrow{R} S^2$, where the latter is the original Hopf map~\cite{Hopf1931}, with $S^n$ being the $n$-sphere. 
Equipped with this composite map, the explicit formula for the Hopf invariant or index $\chi$ is given by~\cite{Pontryagin1941,Ren2007,Wilczek1983,Whitehead1947,Moore2008}
    \begin{align}
        \chi &
        = -\frac{1}{4\pi^2}\int_{T^3}\mathcal{A}\wedge\mathcal{F}
        = - \frac{1}{8\pi^2} \int_{T^3} d^3k \, \epsilon^{\mu\nu\rho}F_{\mu\nu} A_\rho, 
        \label{eq:hermitian-hopf-index}
    \end{align}
where $A_\mu = i\braket{z}{\del_\mu z}$ is the Berry connection and $F_{\mu\nu} = \del_\mu A_\nu - \del_\nu A_\mu=i \left( \braket{\del_\mu z}{\del_\nu z} - \braket{\del_\nu z}{\del_\mu z} \right)$ the Berry curvature. In the case of the Hermitian model~\eqref{eq:mrw-hamiltonian} the Hopf invariant $\chi$~\eqref{eq:hermitian-hopf-index} is the same for both the valence (lower) band and conduction (upper) band, since it is quadratic in the Berry connection $A$.
As a function of $m$ the conventional Hopf index $\chi$ takes on the following values~\cite{Moore2008,He2020} 
\begin{align}
    \chi = \begin{cases}
        -2, & \abs{m} < 1 , \\
        1, & 1 < \abs{m} < 3 , \\
        0, & \abs{m} > 3 .
    \end{cases}
    \label{eq:hermitian-hopf-invariant-values}
\end{align}
where in particular at the transition points $\abs{m}\in\{1,3\}$ the spectrum is gapless and no Hopf invariant can be defined. 
We note that the strong invariant, the Hopf index $\chi \in \mathbb{Z}$~\eqref{eq:hermitian-hopf-index} here, is only integer-quantized provided that the weak invariants, the 2D sub-tori Chern numbers $C_{xy},C_{yz},C_{zx} \in \mathbb{Z}$ given by $2\pi C_{ij} = \int_{T^2_{ij}} F_{ij}$ here, vanish~\cite{Moore2008,Kennedy2016}, which is satisfied for the MRW model~\eqref{eq:mrw-hamiltonian}.
Otherwise, the Hopf index takes values in a finite group, only defined modulo $\mathbb{Z}_{2\gcd(C_{xy},C_{yz},C_{zx})}$~\cite{Kennedy2016}. 
We note that the Hopf invariant $\chi\in\mathbb{Z}$~\eqref{eq:hermitian-hopf-index} is also known as the BZ-integral over the Abelian $U(1)$ Chern-Simons 3-form~\cite{Wilczek1983,Moore2008}. 
The underlying Hopf map $S^3\xrightarrow{R}S^2$ further carries the geometric interpretation of~\cite{Wilczek1983}: the preimages of two generic points on the target $S^2$ are two circles in $S^3$ whose linking number equals the Hopf invariant $\chi$~\cite{Wilczek1983,Whitehead1947,Ren2007,Pontryagin1941}.

In the following, we construct a bosonic Bogoliubov-de~Gennes (BBdG) or symplectic Hopf insulator model, which reduces exactly to the known Hermitian Hopf insulator model~\eqref{eq:mrw-hamiltonian} in a certain parameter limit, and whose classification is \emph{a priori} distinct from all the generalizations mentioned above.

\section{Symplectic Hopf insulator model}
\label{sec:symplectic-hopf-model}
\begin{figure*}[bt]
     \centering
    \hspace*{-.3cm}  %
    \includegraphics[width=2.05\columnwidth]{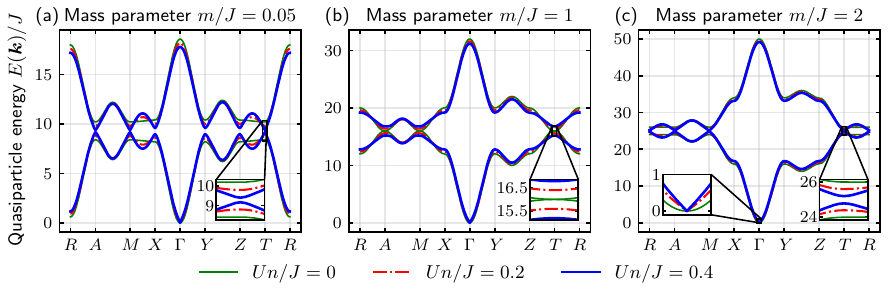}
    \caption{
        BBdG Quasiparticle energy spectrum $E(\kv)$ of the symplectic Hopf insulator model~\eqref{eq:bog-hamiltonian-main}
        for different values of the mass parameter $m/J\in\{0.05,1,2\}$ (a)-(c) and interaction strength $Un/J\in \{0,0.2,0.4\}$ (colored lines) along a high-symmetry path, $R$--$A$--$M$--$X$--$\Gamma$--$Y$--$Z$--$T$--$R$, in the first Brillouin zone. 
        The insets around $T$ in (a) and (c) show how the Hopf-topology-preserving energy gap stays open but shrinks with increasing interaction strength $Un/J$ while in (b) at the critical point $m/J=1$ (gapless for $Un/J=0$) a topologically trivial gap opens for $Un/J>0$~(cf.~\cref{fig:phase-diagram-hopf-bdg}).
        The inset around the $\Gamma$ point in (c) shows the emerging linear Goldstone mode for increasing $Un/J$.
        The high-symmetry points are defined as $R=(k_x/a,k_y/a,k_z/a)=(\pi,\pi,\pi)$, $A=(\pi,0,\pi)$, $M=(\pi,\pi,0)$, $X=(\pi,0,0)$, $\Gamma=(0,0,0)$, $Y=(0,\pi,0)$, $Z=(0,0,\pi)$ and $T=(0,\pi,\pi)$ with the lattice constant $a$.
        }
    \label{fig:hopf-bbdg-model-spectrum}
\end{figure*}
Our construction of a concrete symplectic Hopf insulator model starts from a Bose-Hubbard model of weakly interacting bosons on a three-dimensional cubic lattice with two sublattice sites $s\in\{A,B\}$ per unit cell, where we assume translational invariance along all three spatial directions, such that quasimomenta $\kv=(k_x,k_y,k_z)$ become good quantum numbers.
The associated Hamiltonian is given by
 \begin{align}
    \hat{H} &= -J\sum_{\kv}\sum_{s,s'}\hat{a}^{\dg}_{\kv s}\,[\mathcal{H}(\kv)]_{ss'}\,\hat{a}_{\kv s'}
    \nonumber \\
    &+ \frac{U}{2N_{\mathrm{uc}}}\sum_{\substack{\kv,\kv{'},\kv^{\prime \prime},s}} 
    \hat{a}^{\dg}_{\kv +\kv^{''},s}\hat{a}^{\dg}_{\kv^{'}-\kv^{''},s}\hat{a}_{\kv^{\prime},s}\hat{a}_{\kv,s},
    \label{eq:hopf-bose-hubbard-hamiltonian}
 \end{align}
where single-particle terms in our model correspond to the MRW Hamiltonian $\mathcal{H}(\kv)$ of \cref{eq:mrw-hamiltonian} with an overall hopping energy scale $J$, a repulsive on-site interaction $U>0$, and $N_{\mathrm{uc}}$ the number of unit cells. 
At zero temperature and weak interactions, a Bose-Einstein condensate can form in the lower MRW band. 
By treating the repulsive on-site interaction $U$ as weak at a filling of $n$ bosons per lattice site via a self-consistent Bogoliubov approximation with quantum fluctuations on top of a Gross-Pitaevskii mean-field solution for the Bose-Einstein condensate~\cite{Furukawa2015,Pethick2008,Dalfovo1999}, we obtain a bosonic Bogoliubov-MRW Hamiltonian, our desired \emph{symplectic Hopf insulator} model, 
\begin{widetext}
    \begin{align}
        \hat{H}^{(B)} &= \sum_{\kv\neq\bm{0}}\sum_{s,s'}\Big[
            \hat{a}^{\dg}_{\kv s}\,[\mathcal{H}(\kv)+\mathcal{H}_1]_{ss'}\,\hat{a}_{\kv s'}
            +\big(\hat{a}^{\dg}_{\kv s}\,[\mathcal{H}_2]_{ss'}\,\hat{a}^{\dg}_{-\kv s'}
            + \mathrm{h.c.}\big)\Big]
            \nonumber\\
            &= \frac{1}{2}\sum_{\kv\neq\bm{0}} \hat{\Psi}^{\dg}_{\kv}\,\mathcal{M}_{\kv}\,\hat{\Psi}_{\kv}
            + \mathrm{const.},
            \quad \text{with} \quad
            \mathcal{M}_{\kv} = \begin{pmatrix}
                \mathcal{H}(\kv)+\mathcal{H}_1 & 2\mathcal{H}_2 \\
                2\mathcal{H}_2^{*} & (\mathcal{H}(-\kv)+\mathcal{H}_1)^{*}
            \end{pmatrix}.
            \label{eq:bog-hamiltonian-main}
        \end{align}
\end{widetext}
This a bosonic Bogoliubov-de~Gennes (BBdG) Hamiltonian with the Nambu spinor $\hat{\Psi}_{\kv} = (\hat{a}_{\kv A}, \hat{a}_{\kv B}, \hat{a}^{\dg}_{-\kv A}, \hat{a}^{\dg}_{-\kv B})^T$. 
The normal block $\mathcal{H}(\kv)+\mathcal{H}_1$ contains the single-particle MRW-Hamiltonian~\eqref{eq:mrw-hamiltonian} and the mean-field interaction shift $\mathcal{H}_1 = 4Un\abs{F_-}^2 - \mu_{\mathrm{eff}}\mathbbm{1}_2$, while the pairing (anomalous) block $\mathcal{H}_2 = Un\,F_-^2$ generates the squeezing that carries the interaction effects through the parameter $Un$, with $F_- = \diag{f_{A,-},f_{B,-}}$ the condensate amplitudes on the two sublattices and $n = N/(2N_{\mathrm{uc}})$ the boson density per lattice site (see~\eqref{eq:supp:block-matrix} in~\cref{app:symplectic-hopf} for the full expressions, including the chemical potential $\mu_{\mathrm{eff}}$).
The limit $Un\to0$ switches off the pairing and recovers the Hermitian
MRW insulator physics of~\cref{sec:hermitian-hopf}. 
This has the standard BBdG form $\mathcal{M}= \bigl(\begin{smallmatrix} K & G \\ G^* & K^* \end{smallmatrix}\bigr)$ with $K$ being Hermitian and $G$ being symmetric~\cite{Colpa1978,Tesfaye2025}. 
Note that~\eqref{eq:bog-hamiltonian-main} also contains a separate $\Psi^{\dagger}_{+} \mathcal{M}_{+}\Psi_{+}$ term for the upper band at the condensation point, $\Psi^{\dagger}_{+} \equiv \colvec{\hat{a}^{\dagger}_{\zv,+} \, \hat{a}_{\zv,+}}$, which is left out for brevity here but is discussed in more detail in~\cref{app:symplectic-hopf}, where also the rest of the model details can be found.

The Hamiltonian~\eqref{eq:bog-hamiltonian-main} can be brought to diagonal form
\begin{align}
     \hat{H}^{(B)}=\sum_{\kv} E_{+}(\kv)\hat{b}^{\dagger}_{\kv,+} \hat{b}_{\kv,+}
            +\sum_{\kv \neq \zv} E_{-}(\kv)\hat{b}^{\dagger}_{\kv,-} \hat{b}_{\kv,-}\! +\text{const.},
    \label{eq:bog-hamiltonian-diagonal}
\end{align}
by a Bogoliubov transformation $\hat{\Psi}_{\kv} = W_{\kv}\hat{\Phi}_{\kv}$ that preserves the
bosonic commutation relations, which makes the Bogoliubov transformation matrix $W_{\kv}$ \emph{paraunitary} with respect to the Krein metric
$\tau_z$~\cite{Colpa1978,Schulz-Baldes2017,Shindou2013a,Tesfaye2025},
\begin{align}
    W_{\kv}^{\dg}\,\tau_z\,W_{\kv} = W_{\kv}\,\tau_z\,W_{\kv}^{\dg} = \tau_z,
    \quad
    W_{\kv}^{\dg}\,\mathcal{M}_{\kv}\,W_{\kv} = \bm{E}_{\kv},
    \label{eq:paraunitary}
\end{align}
where $\tau_i = \sigma_i\otimes\mathbbm{1}_2$ with the standard Pauli matrices $\sigma_i$ 
and $\bm{E}_{\kv} = \diag{E_+(\kv),E_-(\kv),E_+(-\kv),E_-(-\kv)}$ the diagonal matrix of Bogoliubov excitation energies. 
For a thermodynamically stable system where the coefficient matrix $\mathcal{M}_{\kv}$ is positive definite~\cite{Colpa1978,Peano2018,Flynn2020}, which we call condition~(S), it is guaranteed that
$\bm{E}_{\kv}$ contains the strictly positive excitation energies $E_\pm(\kv) > 0$. 
When in addition the Bogoliubov spectrum has a uniform gap away from condensation point, [condition~(G)], the particle band defines a continuous map into the Krein-positive projective
domain. 
As a terminological remark, ``symplectic'', follows the BBdG literature~\cite{Peano2016a,Engelhardt2015} as the BBdG Hamiltonian~\eqref{eq:bog-hamiltonian-main} may be equivalently diagonalized via symplectic transformation matrix $\tilde{W}\in \mathrm{Sp}(2N,\mathbb{R})$ when represented in the quadrature representation~\cite{Ferraro2005,Tesfaye2025}. 
This should not be confused with the fact that the Berry curvature defines a symplectic K\"{a}hler 2-form~\cite{Mera2021,Nakahara2003}. 
For a full review of bosonic Bogoliubov theory and its Krein-space structure, see~\cref{app:review-Bogoliubov-Hamiltonian-Indefinite-Inner-Product} or the SM of~\cite{Tesfaye2025}.

In \cref{fig:hopf-bbdg-model-spectrum} we depict the particle-sector Bogoliubov energy spectrum 
$E_{n}(\kv)$~\eqref{eq:bog-hamiltonian-diagonal} along a high-symmetry path in the first Brillouin zone (FBZ) for representative mass terms and interaction strengths with insets showing the direct gap at the high-symmetry point $T$ and the Goldstone branch at $\Gamma$. 
In the topological regime, panels (a) $m/J=0.05$ and (c) $m/J=2$, we see direct gap that
protects the Hopf topology, stays open for every $Un/J$ shown, and shrinks only
moderately as the interactions $Un$ grow. 
At the non-interacting critical mass, panel (b) $m/J=1$, the spectrum is gapless at $Un=0$ and
a gap reopens at finite $Un$, although this gap is topologically trivial, as we will see in \cref{fig:phase-diagram-hopf-bdg}. 
As expected for a weakly interacting Bose gas~\cite{Pethick2008,Dalfovo1999}, near the condensation point $\kv_0 = \bm{0}$, inset in (c), the interactions convert the quadratic band minimum into a linear Goldstone (phonon) branch, $E_-(\kv) \to c(Un)\,\abs{\kv}$ as
$\kv \to \bm{0}$, whose sound speed $c$ grows with $Un$.

\subsection{Symplectic Hopf invariant}
\label{subsec:symplectic-hopf-invariant}
The Bogoliubov modes $\ket{w_n(\kv)}$, the columns of the paraunitary Bogoliubov transformation $W_{\kv}$,
play the role of the Bloch states akin to single-particle wave functions in the conventional Hermitian case~\cite{Peano2016a,Tesfaye2025}.
We stress that $\ket{w_n(\kv)}$ denotes such a single-particle Bogoliubov mode, a coefficient vector in the four-dimensional Nambu space, and is not to be confused with a many-body Fock state such as the multi-mode-squeezed-vacuum ground state of the BBdG Hamiltonian~\eqref{eq:bog-hamiltonian-main}.
These Bogoliubov modes $\ket{w_n(\kv)}$ solve the eigenvalue problem
\begin{align}
    &D_{\kv}\,\ket{w_n(\kv)} = \tilde{E}_n(\kv)\,\ket{w_n(\kv)}, \\
    &D_{\kv} \equiv  \tau_z\mathcal{M}_{\kv}, \  \bra{w_m}\tau_z\ket{w_n} = s_n\,\delta_{mn},
    \label{eq:dyn-matrix-main}
\end{align}
associated to the \emph{pseudo}-Hermitian dynamical matrix $D_{\kv} = \tau_z D_{\kv}^{\dg}\tau_z$~\cite{Peano2016a,Lein2019,Schulz-Baldes2017} which also admits the particle-hole symmetry (PHS) or constraint~\cite{Lein2019}, $D_{\kv} = -\tau_x D_{-\kv}^* \tau_x$. 
Here, Krein sign $s_n = \pm1$ fixes the mode normalization, and $\tilde{E}_n = (\tau_z \bm{E}_{\kv})_{nn} = s_n E_n$ is positive for the particle modes ($s_n = +1$) and negative for the hole modes ($s_n = -1$)~\cite{Shindou2013a,Tesfaye2025}. 
To obtain the correct Bogoliubov quasiparticle energies $E_n(\kv)$, it suffices to (numerically) diagonalize the pseudo-Hermitian dynamical matrix $D_{\kv}$, however, we stress that its eigenstates do \emph{not} automatically adhere to the paraunitary structure of~\eqref{eq:paraunitary}. To obtain the correct paraunitary Bogoliubov transformations $W_{\kv}$, we utilize the method of Colpa~\cite{Colpa1978,Shindou2013a}. 

We focus on the upper particle band with $\ket{w_+(\kv)} \equiv \ket{w_+} = (u, v)^T$ with particle and
hole blocks $u, v \in \mathbb{C}^2$ and $s_+ = +1$, so that Krein positivity reads
$\bra{w_+}\tau_z\ket{w_+} = \abs{u}^2 - \abs{v}^2 = 1 > 0$ (we suppress the parameter dependence on $\kv$ in the following).
The associated symplectic Berry connection $\mathcal{A}_\mu$ and Berry curvature $\mathcal{F}_{\mu \nu}$ are~\cite{Shindou2013a,Furukawa2015,Tesfaye2025} given by
\begin{align}
    \mathcal{A}_\mu = i\,\bra{w_+}\tau_z\ket{\del_\mu w_+},
    \quad
    \mathcal{F}_{\mu\nu} = \del_\mu\mathcal{A}_\nu - \del_\nu\mathcal{A}_\mu, 
    \label{eq:symp-berry}
\end{align} 
which under a $U(1)$ gauge transformation $\ket{w_+} \to e^{i\theta(\kv)}\ket{w_+}$ transform as
$\mathcal{A} \to \mathcal{A} - d\theta$, confirming a genuine $U(1)$ gauge field whose curvature periods
$[\mathcal{F}/2\pi] \in H^2(T^3;\mathbb{Z}) \cong \mathbb{Z}^3$ are the three weak symplectic Chern numbers $C_{yz}, C_{zx}, C_{xy}$ (\cref{app:quantization}).

The particle band defines a map $f : T^3 \to \mathcal{L}^+_2$ into the Krein-positive projective
domain $\mathcal{L}^+_2$, where $\mathcal{L}^+_n$ is given by 
\begin{align}
    \mathcal{L}^+_n = \{[\psi] \in \mathbb{CP}^{2n-1}\! :\! \bra{\psi}\tau_z\ket{\psi} > 0\} \cong
    \frac{U(n,n)}{U(1)\times U(n-1,n)}.
    \label{eq:krein-positive-projective-domain}
\end{align}
which is an open non-compact subset of $\mathbb{CP}^{2n-1}$.
Here, $U(n,m)$ is the indefinite unitary group of signature $(n,m)$, which preserves the indefinite inner product $\bra{\psi}\diag{1_n,-1_m}\ket{\psi}$, and $\mathbb{CP}^{n}$ is the complex projective space of complex dimension $n$.

For general mode number $n$ (with $u,v\in\mathbb{C}^n$), the linear homotopy $r_t : [(u,v)] \mapsto [(u,(1-t)v)]$ with $t\in [0,1]$ stays within $\mathcal{L}^+_n$, since $\abs{u}^2 - (1-t)^2\abs{v}^2 \geq \abs{u}^2 - \abs{v}^2 > 0$, and thereby retracts $\mathcal{L}^+_n$ onto $\mathbb{CP}^{n-1}$ for every $n$ (\cref{app:subsec:quant-classifying}); for $n=2$ this target is $\mathbb{CP}^1$. 
Thus, we have $\mathcal{L}^+_2 \simeq \mathbb{CP}^1 \cong S^2$, so that $\pi_3(\mathcal{L}^+_2) \cong \pi_3(S^2) = \mathbb{Z}$~\footnote{We write $\cong$ for a homeomorphism of spaces or, equivalently, an isomorphism of the associated groups and cohomology, $\simeq$ for a homotopy equivalence of spaces (and, applied to maps, for a homotopy) while retractions are specified in words. Note that a strong deformation retraction is stronger than, and in particular implies, a homotopy equivalence.}.
Hence, a symplectic Hopf invariant exists if and only if the unit cell hosts exactly two
bosonic modes. 
Indeed, $\mathcal{L}^+_1$ is contractible ($\pi_3 = 0$), 
and for $n \geq 3$ modes the fibration $S^1 \to S^{2n-1} \to \mathbb{CP}^{n-1}$ gives $\pi_3(\mathbb{CP}^{n-1}) = 0$, whereas the 
weak symplectic Chern numbers persist, $\pi_2(\mathbb{CP}^{n-1}) = \mathbb{Z}$, for all $n \geq 2$.
This is the bosonic counterpart of the non-stable, delicate character of the fermionic Hopf insulator, which likewise requires exactly two bands~\cite{Kennedy2014,Nelson2021}.
More details are given in~\cref{app:quantization}.

Equipped with the symplectic Berry connection and curvature~\eqref{eq:symp-berry}, the symplectic Hopf invariant can be defined as 
\begin{align}
    \chi = -\frac{1}{8\pi^2}\int_{T^3} d^3k\,\epsilon^{\mu\nu\rho}\mathcal{F}_{\mu\nu}\mathcal{A}_\rho.
    \label{eq:symplectic-hopf-invariant}
\end{align}
As in the conventional Hermitian case~\cite{Moore2008,Kennedy2016}, the Whitehead integral
formula~\eqref{eq:symplectic-hopf-invariant} yields an integer $\chi \in \mathbb{Z}$ only if
$\mathcal{F}$ is (globally) exact, equivalently if the weak symplectic Chern numbers vanish, condition~(W), $C_{yz} = C_{zx} = C_{xy} = 0$.
We show in~\cref{app:quantization} that this condition~(W) holds on every stable gapped component that
is connected to the non-interacting regime $Un = 0$, where $f$ reduces to the standard map
$T^3 \xrightarrow{z} S^3 \to S^2$ and $H^2(S^3;\mathbb{Z}) = 0$ forces all three Chern numbers to vanish.

Finally, note that the BBdG particle-hole structure $\mathcal{M}_{\kv} = \tau_x \mathcal{M}_{-\kv}^{*}\tau_x$
gives that the symplectic Hopf invariant of the hole band is the negative of that of the particle band, 
$\chi_{\mathrm{hole}} = -\chi_{\mathrm{particle}}$, which is different from the $d=2$ symplectic
Chern case~\cite{Shindou2013a,Furukawa2015} where particle and hole bands carry equal Chern numbers provided that the system is thermodynamically stable~(see~\cref{app:quantization} for details).

\subsection{Phase diagram}
\label{subsec:Phase-Diagram}
\begin{figure}[tb!]
    \centering
    \includegraphics[width=\columnwidth]{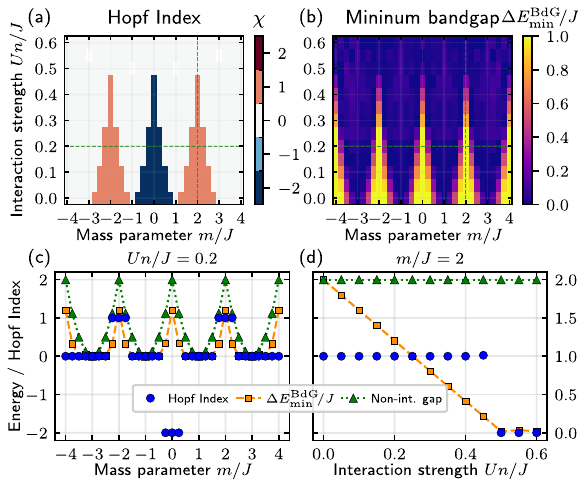}
    \caption{
    (a) Phase diagram of the symplectic Hopf insulator model~\eqref{eq:bog-hamiltonian-main}.
    Symplectic Hopf index $\chi$~\eqref{eq:symplectic-hopf-invariant} of the upper Bogoliubov particle band as a function of the mass parameter $m$ and interaction strength $Un$.
    (b) Minimum Bogoliubov-de~Gennes (BdG) quasiparticle energy band gap $\Delta E^\mathrm{BdG}_\mathrm{min}$ over the same parameter range as in (a).
    Gap closings where $\Delta E^\mathrm{BdG}_\mathrm{min}\to 0$ indicate topological phase transitions between regions of distinct symplectic Hopf invariants. 
    Slices through the phase diagram above at fixed interaction strength $Un/J=0.2$ (c) and fixed mass parameter $m/J=2$ (d), as indicated by the green-colored dashed lines in (a) and (b).
    The symplectic Hopf index $\chi$ (blue circles) and minimum BdG band gap $\Delta E^\mathrm{BdG}_\mathrm{min}$ (orange squares) are shown as functions of the mass parameter $m$ (c) and interaction strength $Un$ (d) while the green triangles depict the non-interacting ($Un/J=0$) bare gap.
    Integer-valued plateaus of the symplectic Hopf invariant $\chi$ coincide with regions where finite BBdG quasiparticle gap is present with gap closings marking topological phase transitions. 
    A $N^3 = 100^3$ quasimomentum grid has been used~(cf.~\cref{app:numerical-hopf-invariant}).
    }
    \label{fig:phase-diagram-hopf-bdg}
\end{figure}

For the upper particle band of symplectic Hopf insulator model of~\eqref{eq:bog-hamiltonian-main}, we numerically compute the symplectic Hopf invariant $\chi$~\eqref{eq:symplectic-hopf-invariant} as function of the mass parameter $m$ and interaction strength $Un$ and depict the results in~\cref{fig:phase-diagram-hopf-bdg}. 
As the integrand of \eqref{eq:symplectic-hopf-invariant} is gauge-dependent, we use an adapted numerical algorithm of~\Ccite{Moore2008,He2020}, which fixes the gauge in Fourier space, to compute $\chi$~(see ~\cref{app:numerical-hopf-invariant} for details). 
As shown in~\cref{fig:phase-diagram-hopf-bdg}(a), we find that, away from the non-interacting MRW-limit $Un = 0$, the symplectic Hopf invariant $\chi$ remains integer-quantized and stable against moderate interaction strengths where the most stable regimes are close to the center of the non-interacting topological windows, $\abs{m}/J=0,2$, where the Hopf topology is robust up to $Un/J \sim 0.45$. 
At $Un = 0$ the topological windows reproduce the bare MRW result
\eqref{eq:hermitian-hopf-invariant-values}, with gap closings occurring at $\abs{m} =1, 3$
marking the transitions. 
The phase boundaries shift moderately with increasing $Un$, as the isotropic pairing terms in~\eqref{eq:bog-hamiltonian-main} progressively narrow the BBdG gap.
The integer-quantized topological regions of stable symplectic Hopf invariants $\chi$ coincide with regions where the minimum Bogoliubov quasiparticle energy gap between the lower and upper particle band, $\Delta E^{\mathrm{BdG}}_{\mathrm{min}} = \min_{\kv}\abs{E_+(\kv) - E_-(\kv)}$, 
stays open over the Brillouin zone, as shown in~\cref{fig:phase-diagram-hopf-bdg}(b). 
In~\cref{fig:phase-diagram-hopf-bdg}(c) and (d), we show one-dimensional slices of the phase diagram at fixed $Un/J = 0.2$ and fixed $m/J = 2$, respectively. 
These results corroborate the symplectic Hopf invariant $\chi$ as the correct invariant for bosonic BdG systems. 

\subsection{Bulk-boundary correspondence and edge states}
\label{subsec:edge-states}
\begin{figure*}[t]
     \centering
    \includegraphics[width=2.1\columnwidth]{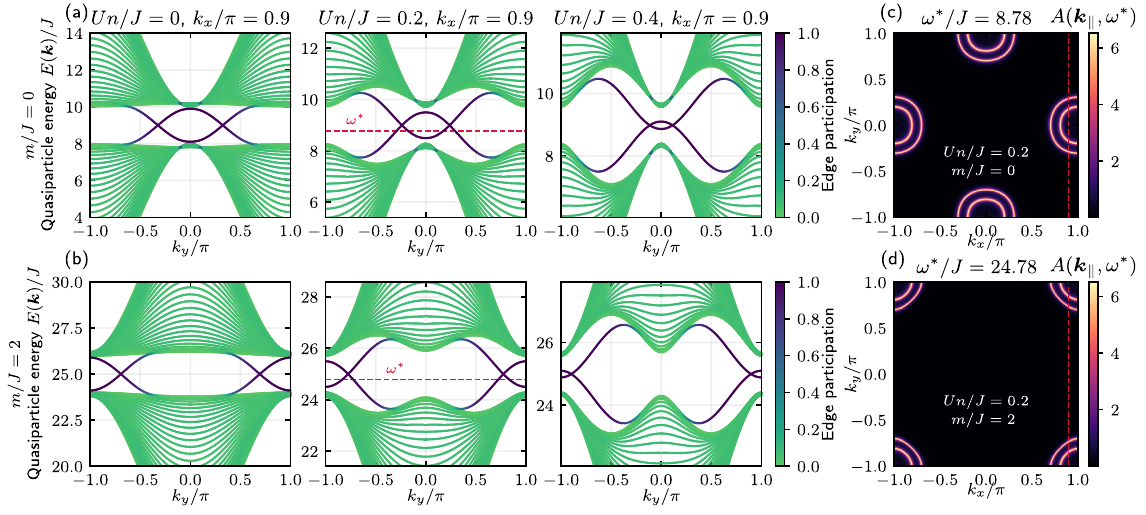}
    \caption{
    (a)-(b) BBdG Quasiparticle energy spectrum $E(\kv)$ of the symplectic Hopf insulator model~\eqref{eq:bog-hamiltonian-main} as function of $k_y$ for fixed $k_x/\pi = 0.9$ for open boundary conditions along the $z$-direction for two mass parameter $m/J=0$ (a) and $m/J=2$ (b) (rows) and different interaction strengths $Un/J\in \{0,0.2,0.4\}$ (columns).
    The color scale indicates the symplectic edge participation $p_j(\kv_\parallel)$~\eqref{eq:edge-participation}, i.e., (the fraction of) the state's probability living on the surface with the correct $\tau_z$-norm with dark (bright) colors indicating localization on the surface (bulk) of the system. 
    The bulk-boundary correspondence in the presence of finite interactions is confirmed by the appearance of topologically protected in-gap surface states in the regimes of non-zero symplectic Hopf indices $\chi$ (cf.~\cref{fig:phase-diagram-hopf-bdg}). 
    (c)-(d) 
    The BBdG surface spectral function $A(\kv_{\parallel},\omega)$~\eqref{eq:surface-spectral-function} with $\kv_{\parallel} = (k_x,k_y)$
    for the interaction strength $Un/J=0.2$ and mass parameters $m/J=0$ (c) and $m/J=2$ (d) at energy $\omega^*/J=8.78$ (c) and $\omega^*/J=24.78$ (d) (see red-colored dashed lines in the second column of (a) and (b)), where the red-colored dashed lines here indicate the $k_x/\pi=0.9$ cut along which the spectra in (a) and (b) are computed.
    The topological surface states are physically observable as the surface spectral function $A(\kv_{\parallel},\omega)$ peaks at the finite energy values $\omega^*$ where they occur.
    We have used $L_z = 32$ sites along the $z$-direction and a Lorentzian broadening factor of $\eta = 0.05 J$ for the surface spectral function $A(\kv_{\parallel},\omega)$ in (c) and (d).
    }
    \label{fig:hopf-bbdg-model-chiral-edge-modes}
\end{figure*}
Now, we study the topologically protected surface states of the symplectic Hopf insulator model. 
Note that in the Hermitian Hopf insulator, the bulk-boundary correspondence is delicate: the Hopf invariant $\chi$ does not protect gapless chiral edge modes as a Chern number would, but it does guarantee that at least $\abs{\chi}$ surface modes appear on any boundary~\cite{Deng2013,Nelson2021}. 
In our BBdG case, we open the cubic lattice along $z$, keeping $\kv_\parallel = (k_x,k_y)$ as good quantum numbers, via the inverse DFT method of~\Ccite{Deng2013}, and diagonalize the resulting bosonic BdG slab para-unitarily~\eqref{eq:dyn-matrix-main} to obtain the BBdG quasiparticle spectrum $E_j(\kv_\parallel)$ and the associated slab Bogoliubov modes $\ket{w_j(\kv_\parallel)}= (u_j, v_j)^T$. 
In addition to this spectrum, we compute the surface localization of the slab modes on the terminating layer sites $S$ of the slab via
\begin{align}
    p_j(\kv_\parallel) 
    = \frac{1}{s_j}\sum_{c\in S}(\abs{u_j(c)}^2 - \abs{v_j(c)}^2)
    \label{eq:edge-participation}
\end{align}
which we call the edge participation of each mode and where $c$ labels the real-space sites of the open slab. 
Here, $s_j = \bra{w_j}\tau_z\ket{w_j} = \pm1$ is the Krein sign, so that $p_j$ is the fraction of the conserved $\tau_z$ norm carried by the surface with $j\in\{1,\dots,2L_z\}$ labeling the slab modes~\cite{Furukawa2015,Shindou2013a}.
Both the slab BBdG quasiparticle energy spectrum $E_j(\kv_\parallel)$ and the corresponding edge participation $p_j(\kv_\parallel)$ are shown in~\cref{fig:hopf-bbdg-model-chiral-edge-modes}(a),(b) as function of $k_y$ at fixed $k_x/\pi=0.9$ for interaction strengths $Un/J \in \{0, 0.2, 0.4\}$
for the topological masses $m/J = 0, 2$ ($\chi = -2, +1$~[\cref{fig:phase-diagram-hopf-bdg}]), respectively. 
We find symplectic topological surface states appear in the gap between the upper and lower quasiparticle energy spectrum at parameter regions where the phase diagram~[\cref{fig:phase-diagram-hopf-bdg}] shows the symplectic Hopf invariant $\chi$ is non-zero.
The edge participation $p_j(\kv_\parallel) \approx 1$ shows that these surface states are indeed localized on the terminating layers of the slab, while the bulk bands with $p_j(\kv_\parallel) \approx 0$ are delocalized over the entire slab. 
The surface states are progressively pushed into the bulk bands as the interaction strength $Un$ is increased, consistent with the narrowing of the bulk gap in~\cref{fig:phase-diagram-hopf-bdg}(b), which, however, might not close at the momentum slice depicted in~\cref{fig:hopf-bbdg-model-chiral-edge-modes}(a),(b) but at other $\kv_\parallel$ values~[cf.~\cref{fig:hopf-bbdg-model-spectrum}]. 

Contrary to the fermionic case where the surface states cross the Fermi energy, here the topological surface states lie at a finite energy $\omega^* > 0$ above the condensate. 
To excite these topological surface states, which are isolated from the bulk modes in momentum and energy, one can utilize a high-frequency Raman pulse to realize so-called edge matter waves~\cite{Furukawa2015}. 
The response of the system to such finite-frequency probes can be analyzed via the boundary-projected surface spectral function $A(\kv_\parallel,\omega)$~\cite{Furukawa2015}, defined as
\begin{align}
    A(\kv_\parallel,\omega) = \sum_j
    w^u_j(\kv_\parallel)\,\delta_\eta(\omega-E_j)
    + w^v_j(-\kv_\parallel)\,\delta_\eta(\omega+E_j),
    \label{eq:surface-spectral-function}
\end{align}
with boundary weights $w^u_j = \sum_{c\in S}\abs{u_j(c)}^2$, $w^v_j = \sum_{c\in S}\abs{v_j(c)}^2$ and a Lorentzian regularization $\delta_\eta(x) = \eta/[\pi(x^2+\eta^2)]$: a particle peak at $+E_j$ with residue $\abs{u_j}^2$ and a hole peak at $-E_j$, evaluated at $-\kv_\parallel$, with residue $\abs{v_j}^2$.  
Here, we depict $A(\kv_\parallel,\omega^*)$ in~\cref{fig:hopf-bbdg-model-chiral-edge-modes}(c),(d), at the energy $\omega^*$ of the surface states in (a)-(b) at $Un/J = 0.2$ and $m/J = 0, 2$ (red-colored dashed line) respectively.
Indeed, we find that the surface states appear as peaked arcs in the surface spectral function $A(\kv_\parallel,\omega^*)$ and are thus measurable via Raman or Bragg spectroscopy~\cite{Stamper-Kurn1999,Stenger1999}. 

Lastly, we note that the protection of these surface states is \emph{delicate}. 
The reason for this is that the condition~(W) forces the bulk weak symplectic Chern numbers on the coordinate sub-tori to vanish, so on any fixed-$k_x$ (or fixed-$k_y$) cut the net number of chiral surface crossings set by the sub-torus Chern number, is zero, and the surface states carry no per-cut chirality. 
This does not render the boundary trivial. In the Hermitian Hopf insulator the surface-localized bands
carry a quantized \emph{faceted} Chern number equal to the bulk Hopf index, $C_{\mathrm{f}} = \chi$,
a boundary anomalous Hall response that is visible only out of equilibrium~\cite{Nelson2021,Lapierre2021},
and we expect the same to hold for the Bogoliubov surface bands. 
The at-least-$\abs{\chi}$ in-gap surface states are thus guaranteed by the Hopf invariant itself and not
by any weak (Chern) invariant, in the same delicate, non-stable sense as in the conventional Hermitian
case~\cite{Deng2013,Nelson2021,Lapierre2021}.

\section{Summary and outlook}
\label{sec:summary-outlook}
We have constructed a symplectic Hopf insulator, via a bosonic Bogoliubov-de~Gennes 
generalization of the Moore-Ran-Wen model~\cite{Moore2008}, starting from a two-sublattice Bose-Hubbard Hamiltonian with weak on-site interactions treated via a self-consistent Bogoliubov approximation including mean-field and quantum fluctuation corrections. 
Using the paraunitary structure of BBdG systems and the symplectic Berry connection and curvature, we have proposed the symplectic Hopf invariant $\chi$ as a topological invariant for three-dimensional BBdG systems.
We have found that the symplectic Hopf invariant $\chi$ is integer-quantized and stable against moderate interaction strengths for our model in various mass parameter regions, which further reproduces the non-interacting Hopf invariant in the $Un = 0$ limit exactly. 

This robustness rests on our central topological result, a strong deformation retraction of the Krein-positive projective domain $\mathcal{L}^+_2$ onto $S^2$, which yields $\pi_3(\mathcal{L}^+_2) = \mathbb{Z}$ and thereby protects an integer-quantized symplectic Hopf invariant $\chi$ for the BBdG particle band.
Opening the system along one direction, we have found that the BBdG quasiparticle spectrum hosts topologically protected in-gap surface states at finite excitation energy, observable via
Bragg or Raman spectroscopy.

Several further directions remain open.
First, it will be interesting to explore whether the delicate multiband route of~\Ccite{Lapierre2021}, that restores the $\mathbb{Z}$ invariant for all $N$-band Hopf insulators ($N\geq 2$) by resolving every band into its own gap (classifying space $SU(N)/U(1)^{N-1}$), carries over to the bosonic Krein-space setting. 
Further, another open task is to come up with pairing-generated BBdG Hopf phases with no Hermitian parent, that is BBdG models of the type~\eqref{eq:bog-hamiltonian-main} where the Hopf topology is entirely carried by the off-diagonal non-particle-number conserving pairing blocks. 
Natural candidates for this are systems with long-range interactions or three-dimensional photonic setups with a nonlinear crystal arrays that are coherently pumped generating the required squeezing~\cite{Notomi2008, Mookherjea2002, Eggleton2011, Dahdah2011, Smirnova2020, Peano2016, Peano2016a, Shi2017}. 
Lastly, building upon the measurement proposal for the symplectic quantum geometry
developed in~\cite{Tesfaye2025}, it seems plausible that both topological optical chirality dichroism~\cite{Jankowski2026} with the corresponding gerbe structure~\cite{Palumbo2019,Jankowski2025} and nonlinear viscoelastic effects~\cite{Jain2026} can be established in BBdG systems for measuring the herein proposed symplectic Hopf invariant $\chi$.

\begin{acknowledgments}
We are grateful to D.-L.~Deng for sharing his code implementation of the algorithm for the computation of the conventional Hopf index developed by J.~Moore, Y.~Ran, and X.-G.~Wen~\cite{Moore2008}. 
We thank Marco Di Liberto for insightful discussions and comments. 
G. P. acknowledges support from national funds by FCT - Fundação para a Ciência e Tecnologia, I.P. in the framework of the project UID/04564/2025, with DOI identifier 10.54499/UID/04564/2025.
I.T.~acknowledges support by the Deutsche Forschungsgemeinschaft (DFG, German Research Foundation) via the Research Unit FOR 5688 (Project No. 521530974) and from the Studienstiftung des deutschen Volkes. 
\end{acknowledgments}

\section*{Data and code availability}
Data management has been handled by pywatson~\footnote{I. Tesfaye,~\href{https://github.com/isaac-tes/pywatson}{PyWatson - A Python scientific project management tool} (2026)}. The data that support the findings of this study are openly available at~\footnote{I. Tesfaye and G. Palumbo, Data for “Symplectic Hopf Insulator: Delicate Topology in Bosonic Bogoliubov-de Gennes Systems,” (v1.0.2)~[Data set],~\href{https://doi.org/10.5281/zenodo.22691589}{Zenodo (2026) doi:10.5281/zenodo.22691589}.}.

\clearpage
\makeatletter
\let\addcontentsline\origaddcontentsline
\makeatother
\appendix
\crefalias{section}{appendix} 

\onecolumngrid

\begin{center}
   \Large\textbf{Appendices}
\end{center}
\twocolumngrid
\tableofcontents
\twocolumngrid

\section{Hermitian Hopf insulator}
\label{app:hermitian-hopf}
We complement the review of the Hermitian Hopf insulator model~\cite{Moore2008,Deng2013,He2020} of~\cref{sec:hermitian-hopf} in the main text by providing the explicit momentum-space components of the Bloch vector $\bm{d}(\kv)$ and the two band eigenstates.
By defining
    \begin{align}
        \ket{u} = \begin{pmatrix} u_1(\kv) \\ u_2(\kv) \end{pmatrix}, \quad
        \begin{aligned}
            u_1(\kv) &= \sin k_x + i \sin k_y, \\
            u_2(\kv) &= \sin k_z + i M(\kv),
        \end{aligned}
    \end{align}
where $M(\kv) = m + \cos k_x + \cos k_y + \cos k_z$,
the two-band Hopf insulator Hamiltonian reads
    \begin{align}
        H(\kv) &= \bm{d}\cdot \bm{\sigma}= \sum_{i=1}^{3} d_i\sigma_i, \nonumber \\
        d_i &= \expval{u}{\sigma_i}= \sum_{a,b=1}^2 u_a^* (\sigma_i)_{ab} u_b,
        \label{app:eq:hermitian-hopf-Hamiltonian}
    \end{align}
with Pauli matrices $\bm{\sigma}=(\sigma_1, \sigma_2, \sigma_3)$ and mass parameter $m$. The explicit momentum-space components of $\bm{d}$ follow with $w\equiv u_1^* u_2 = \re{w}+i\,\im{w}$,
    \begin{align}
        \re{w} &=\sin k_x\sin k_z+M(\kv)\sin k_y, \nonumber \\
        \im{w} &=M(\kv)\sin k_x-\sin k_y\sin k_z,
    \end{align}
so that, using the shorthand $M \equiv M(\kv)$,
\begin{align}
        d_1 &= 2\re{w} = 2[\sin k_x\sin k_z+M\sin k_y], \nonumber \\
        d_2 &= 2\im{w} = 2[M\sin k_x-\sin k_y\sin k_z], \nonumber \\
        d_3 &= \abs{u_1}^2 - \abs{u_2}^2 \nonumber \\
        &= \sin^2 k_x+\sin^2 k_y-\sin^2 k_z-M^2.
        \label{app:eq:hermitian-hopf-d-vectors}
    \end{align}
The eigenvalues are $E_\pm = \pm\abs{\bm{d}}$, with $\abs{\bm{d}} = \sqrt{d_1^2 + d_2^2 + d_3^2} = \abs{u_1}^2 + \abs{u_2}^2 = \braket{u}{u}$. We drop the explicit $\kv$-dependence in the following for brevity.
In terms of the normalized spinor $\ket{z} = \ket{u}/\sqrt{\braket{u}{u}}$ of~\cref{sec:hermitian-hopf}, the Hamiltonian reads $H = \abs{\bm{d}}(2\projector{z} - \mathbbm{1}_2)$, so that $\ket{z}$ is the upper band, $H\ket{z} = +\abs{\bm{d}}\ket{z}$, and the lower band is $\ket{\tilde z} = (-z_2^*, z_1^*)^T$ with reversed texture $\tilde{\bm{R}} = -\bm{R}$.
With the Berry connection $A_\mu = i\braket{z}{\del_\mu z}$ and curvature $F_{\mu\nu} = \del_\mu A_\nu - \del_\nu A_\mu$ of~\cref{sec:hermitian-hopf}, the Hopf invariant reads~\cite{Pontryagin1941,Ren2007,Wilczek1983,Whitehead1947,Moore2008,Deng2013,He2020}
    \begin{align}
        \chi = -\frac{1}{4\pi^2} \int A \wedge F
         = -\frac{1}{8\pi^2} \int_{\text{BZ}} d^3k \, \epsilon^{\mu\nu\rho}F_{\mu\nu} A_\rho.
        \label{app:eq:hopf-invariant}
    \end{align}
For the model~\eqref{app:eq:hermitian-hopf-Hamiltonian} it takes the integer values quoted in~\eqref{eq:hermitian-hopf-invariant-values}~\cite{Moore2008}.

\section{Numerical implementation of the Hopf invariant}
\label{app:numerical-hopf-invariant}
Since more complicated generalizations of the Hopf insulator model do not allow for an analytical computation of the Hopf invariant~\cite{He2020}, we briefly review the numerical implementation of the algorithm to compute the Hopf invariant as developed by J.~Moore, Y.~Ran, and X.-G.~Wen in~\cite{Moore2008} and further discussed in~\cite{He2020}. 
We use a gauge invariant spectral form which converges significantly faster than the original finite-difference implementation of~\cite{Moore2008} and can be applied to both the Hermitian and symplectic Hopf insulator models. 
We first review the algorithm for the Hermitian case and then discuss the modifications required for the symplectic Hopf insulator model afterwards.

First, we numerically discretize the 3D Brillouin zone $T^3$ into a grid of $N^3$ points, where $N$ is the number of discretization points along each momentum direction (which we take to be the same for simplicity). 
By numerically solving for the spectrum $H(\kv)\ket{u_n(\kv)} = E_n(\kv)\ket{u_n(\kv)}$
of $H(\kv)$~\eqref{app:eq:hermitian-hopf-Hamiltonian}, we can compute the projectors $P_n(k)=\projector{u_n(\kv)}$. 
We then compute the Berry curvature $F_{\mu\nu}$ and its associated dual current one-form $B_\rho$ via
    \begin{align}
        F_{\mu\nu} = i\, \mathrm{Tr}\left( P [\del_\mu P, \del_\nu P]\right), \quad 
        B_{\rho} = \frac{1}{2} \epsilon_{\rho\mu\nu} F_{\mu\nu},
    \end{align}
where have omitted the band index $n$ for simplicity and the $\kv$-dependence for brevity.
Instead of evaluating the projector derivatives $\del_\mu P$ by finite differences on the grid, we use the exact spectral representation of the Berry curvature of band $n$~\cite{Resta2011}
    \begin{align}
        F_{\mu\nu} &= - 2\,\mathrm{Im}\left[\sum_{m\neq n} \frac{\matel{u_n}{\del_\mu H}{u_m}\matel{u_m}{\del_\nu H}{u_n}}{(E_n - E_m)^2} \right] \nonumber \\
        &= -2\,\mathrm{Im} \left[ \sum_{m\neq n}
         \frac{\mathrm{Tr} \left(P_n(\del_\mu H)P_m(\del_\nu H)\right)}{(E_n-E_m)^2} \right],
        \label{app:eq:spectral-berry-curvature}
    \end{align}
which follows from the Hellmann-Feynman theorem. At each isolated grid point $\kv$ it needs only the eigenstates and eigenvalues $E_n(\kv)$ together with the analytical momentum derivatives $\del_\mu H(\kv)$ of the Bloch Hamiltonian~\eqref{app:eq:hermitian-hopf-Hamiltonian}, which are known in closed form for the tight-binding model so that no numerical differentiation of the Bloch states across the grid is required and~\eqref{app:eq:spectral-berry-curvature} is exact. 

Next, to circumvent the issue of gauge-dependence of the Berry connection $A_\mu$ in the Hopf density~\eqref{app:eq:hopf-invariant}, we solve for $A_\mu$ in Fourier space by using both curl equation $\nabla \times \bm{A} = \bm{B}$ ($\Leftrightarrow dA = F$) and a specified gauge choice $\nabla \cdot\bm{A}=0$ (Coulomb gauge).
To this end, we first transform the dual current $B_\rho$ to Fourier space via a discrete Fourier transform~\cite{Deng2013,He2020} (or better FFT) yielding 
    \begin{align}
        B_\rho(\bm{r}) = \sum_{k} e^{-i\kv\cdot\bm{r}} B_\rho(\kv),
    \end{align}
where $\bm{r}$ is the real-space coordinate corresponding to the Fourier space momentum $\kv$.
The curl equation then becomes $(-i\bm{r}) \times \bm{A} = \bm{B}$ and the gauge condition becomes $-i\bm{r} \cdot \bm{A} = 0$. 
By this, it follows from $(-i\bm{r}) \times[(-i\bm{r}) \times \bm{A}] = (-i\bm{r}) \times \bm{B}$ that 
    \begin{align}
        \bm{A}(\bm{r}) =-i \frac{\bm{r} \times \bm{B}(\bm{r})}{r^2}.
    \end{align}
Since the curvature $\bm{B}$ now comes from analytic momentum derivatives instead of a finite-difference stencil, the continuum multiplier $-i\bm{r}$ is used directly, with no lattice correction to the Fourier derivative required.
Having obtained $A_\mu(\bm{r})$ we can then compute the Hopf invariant $\chi$ in (real) Fourier space as
    \begin{align}
        \chi &= -\frac{1}{4\pi^2} \int_{\text{BZ}} d^3k \, \bm{A}(\kv) \cdot \bm{B}(\kv) \nonumber \\
        &= - \frac{(2\pi)^3}{N^3} \frac{1}{4\pi^2} \sum_{\bm{r}} \bm{B}(-\bm{r})\cdot \bm{A}(\bm{r}).
    \end{align}

The same algorithm above is used to compute the symplectic Hopf invariant~\eqref{eq:symplectic-hopf-invariant} of the BBdG model. 
The only change is that the standard Berry curvature is replaced by its symplectic counterpart $\mathcal{F}_{\mu\nu} = -2\,\im{\eta_{\mu\nu}}$, obtained from the symplectic quantum geometric tensor of band $n$~\cite{Tesfaye2025}
    \begin{align}
        \eta_{\mu\nu} &= s_n\sum_{m\neq n} s_m\,
        \frac{\matel{w_n}{\del_\mu\mathcal{M}}{w_m}\matel{w_m}{\del_\nu\mathcal{M}}{w_n}}{(\tilde{E}_n - \tilde{E}_m)^2} \nonumber \\
        &= \sum_{\substack{m=1\\m\neq n}}^{2N}\frac{
            \mathrm{Tr}(P_n(\del_\mu D)P_m(\del_\nu D))}{(\tilde{E}_n - \tilde{E}_m)^2},
        \label{app:eq:sqgt-spectral}
    \end{align}
the direct analog of~\eqref{app:eq:spectral-berry-curvature}. 
The ordinary projectors are replaced by the pseudo-Hermitian projectors of the pseudo-Hermitian dynamical matrix $D_{\kv} = \tau_z\mathcal{M}_{\kv}$~\eqref{eq:dyn-matrix-main}, the analytic velocity matrices $\del_\mu H$ by $\del_\mu\mathcal{M}_{\kv}$, and the energies $E_n$ by the signed dynamical eigenvalues $\tilde{E}_n = s_n E_n$, with the Krein signs $s_n s_m$ following from the pseudo-Hermitian normalization $\bra{w_m}\tau_z\ket{w_n} = s_n\delta_{mn}$. The projectors are obtained from $P_n =W \Gamma^n \tau_z W^\dg \tau_z$ with the paraunitary Bogoliubov transformation $W$ and the diagonal matrix $\Gamma^n$ with $(\Gamma^n)_{mm} = \delta_{mn}$, as discussed in~\cite{Tesfaye2025}.
The subsequent gauge fixing and Fourier reconstruction of $\mathcal{A}$ from $\mathcal{F}$ are unchanged and work in the same way as for the Hermitian case.

\section{Symplectic Hopf insulator model}
\label{app:symplectic-hopf}

Here, we provide details on the symplectic Hopf insulator model considered in this work.
The symplectic Hopf insulator model is derived from a self-consistent mean-field GPE and Bogoliubov approximation for a weakly interacting Bose-Einstein condensate (BEC), starting from a Bose-Hubbard generalization of the Moore-Ran-Wen (MRW) Hopf model~\cite{Moore2008}.
The following derivation, which is mainly based on ~\cite{Tesfaye2025,Furukawa2015}, works for generic two-band models in momentum space.

    The starting point is a Bose-Hubbard model on a three-dimensional (real-space) lattice with two sublattice sites,
     \begin{align}
        \hat{H}=\sum_{\ell \ell',ss'}h_{\ell s,\ell ' s'}\hat{a}^{\dagger}_{\ell s}\hat{a}_{\ell' s'}
        +\frac{U}{2}\sum_{\ell,s} \hat{a}^{\dagger}_{\ell s}\hat{a}^{\dagger}_{\ell s}\hat{a}_{\ell s}\hat{a}_{\ell s},
         \label{eq:supp:BH-Hamiltonian}
    \end{align}
    where the elementary unit cells are labeled by $\ell$, the sublattice sites by $s \in \{A,B\}$.
    The non-interacting (quadratic) part is governed by the matrix elements $h_{\ell s,\ell ' s'}$ 
    in position space and the interacting part by the on-site interaction strength $U$.
    
    Rotating this Bose-Hubbard Hamiltonian~\eqref{eq:supp:BH-Hamiltonian} into quasimomentum space, yields
            \begin{align}
                \hat{H}=&\sum_{\kv} \sum_{s,s'} \hat{a}^{\dagger}_{\kv, s}
                 \mathcal{H}_{s,s'}(\kv) \hat{a}_{\kv, s'} \nonumber \\
                &+\frac{U}{2 N_{\text{uc}}} \sum_{\kv,\kv',\kv''}\sum_{s}
                \hat{a}^{\dagger}_{\kv+\kv'' , s}\hat{a}^{\dagger}_{\kv'-\kv'' , s}
                \hat{a}_{\kv , s}\hat{a}_{\kv' , s}.
                \label{eq:supp:Ham-Full-2band-comb-momentum}
            \end{align}
    Here, we have introduced the bosonic operators transformed into quasimomentum space via 
        \begin{align}
            \hat{a}_{\kv , s} &= \sum_{\ell } \braket{\kv s}{\ell s} \hat{a}_{\ell s} = \sum_{\ell } \frac{e^{i \kv \cdot \rv_{\ell s}}}{\sqrt{N_{\text{uc}}}}
            \hat{a}_{\ell s}, \nonumber \\
        \hat{a}_{\ell s} &= \sum_{\kv } \braket{\ell s}{\kv s} \hat{a}_{\kv , s} = \sum_{\kv } \frac{e^{-i \kv \cdot \rv_{\ell s}}}{\sqrt{N_{\text{uc}}}}
            \hat{a}_{\kv , s},
            \label{eq:supp:Fourier-transf-ops}
        \end{align}
    where the second equation represents the inverse Fourier transform and $\ket{\kv s}$ denotes the quasimomentum-space basis state at sublattice site $s$ with $N_{\text{uc}}$ being the number of unit cells in the system.
    
    Let us first focus on the non-interacting part, the first summand of ~\eqref{eq:supp:Ham-Full-2band-comb-momentum}, which is governed by the quasimomentum-space Hamiltonian $\mathcal{H}_{s s'}(\kv)$, which can be expressed in terms of the Pauli matrices as
    \begin{align}
        \mathcal{H}_{s s'}(\kv)& \equiv \sum_{\ell \ell '}  h_{\ell s,\ell ' s'}
        e^{-i(\rv_{\ell s}-\rv_{\ell' s'})\cdot \kv} \nonumber \\
           &=\left(h_0(\kv) \mathbbm{1} +\bm{h}(\kv)\cdot \bm{\hat{\sigma}}\right)_{s s'},
       \label{eq:supp:Hk-Pauli-exp}
   \end{align}
   where $\bm{\hat{\sigma}} = (\sigma_x,\sigma_y,\sigma_z)^T$ is the vector of Pauli matrices.
   At each quasimomentum point, $\kv$, the Hamiltonian $\mathcal{H}_{s s'}(\kv)$ describes an operator acting on a two-dimensional 
   Hilbert space, whose elements can be represented on the Bloch sphere. 
   The north and south pole are denoted by $\ket{\kv A}$ and $\ket{\kv B}$ respectively. 
   Due to the non-interacting nature of the system, the model can be further diagonalized by rotating into the eigenbasis $\ket{\kv \pm}$,
    \begin{align}
        \ket{\kv -}&=\sin(\theta_{\kv}/2) \ket{\kv A} -\cos(\theta_{\kv}/2)e^{i \varphi_{\kv}} \ket{\kv B}, \nonumber \\
        \ket{\kv +} &=\cos(\theta_{\kv}/2) \ket{\kv A} +\sin(\theta_{\kv}/2)e^{i \varphi_{\kv}} \ket{\kv B}.
        \label{eq:supp:kvec-plusmin-param}
    \end{align}
    which reside at $\pm \hat{\bm{h}}(\kv)$ on the Bloch sphere, with
    $\hat{\bm{h}}(\kv)=\bm{h}(\kv)/|\bm{h}(\kv)|=(\cos(\varphi_{\kv}) \sin(\theta_{\kv}),\sin(\varphi_{\kv})\sin(\theta_{\kv}),\cos(\theta_{\kv}))^T$.
    The corresponding eigenenergies $\epsilon_{\pm}(\kv)$ of $\ket{\kv \pm}$ are given by $\epsilon_{\pm}(\kv)=h_0(\kv)\pm \abs{\bm{h}(\kv)}$
    which define the band structure of the lattice. 
    
    In our case the non-interacting part of the Hamiltonian $\mathcal{H}(\kv)$ is given by the MRW Hopf model~\eqref{app:eq:hermitian-hopf-Hamiltonian}-\eqref{app:eq:hermitian-hopf-d-vectors}, which is known to have a non-trivial Hopf invariant $\chi$~\eqref{app:eq:hopf-invariant} for certain values of the mass parameter $m$~\eqref{eq:hermitian-hopf-invariant-values}.
    For convenience, we associate a natural energy scale $J$ to the non-interacting part of the Hamiltonian $\mathcal{H}(\kv)$, which can be thought of as a hopping amplitude in the underlying Bose-Hubbard model.

    Now, note that an ideal Bose gas at zero temperature [with $U=0$ in~\eqref{eq:supp:BH-Hamiltonian}]
    forms a perfect Bose-Einstein condensate (BEC)
    characterized by all bosons condensing into the lowest energy eigenstate $\ket{\kv_0,-}$ at some quasimomentum $\kv_0$ 
    in the lower band $\epsilon_{-}(\kv)$, 
    with the ground state given by 
    \begin{align}
        \ket{\psi_B}= \frac{(\hat{a}^{\dagger}_{\kv_0 -})^{N}}{\sqrt{N!}} \vac,
    \end{align}
    where we assume a non-degenerate minimum at $\kv_0$ for simplicity.
    Here, $N$ shall denote the total number of bosons in the system and $\hat{a}^{\dagger}_{\kv,\pm}$ the creation operator
    for a boson at quasimomentum $\kv$ in the lower or upper band, respectively.

    Having discussed the non-interacting part, we turn to the Bogoliubov theory, which accounts for additional 
    quantum (and thermal) fluctuations on top of a mean-field solution for a weakly interacting bosonic system~\cite{Ueda2010,Pethick2008,Dalfovo1999}.
    For this, we first obtain a mean-field solution via a minimization of an associated Gross-Pitaevskii (GP) energy functional.
    The stationary solutions gained by minimizing the GP energy functional will then yield the mean-field ground state, which
    is assumed to be macroscopically occupied.
    Then, we will describe quantum fluctuations on top of the condensate, by considering
    non-condensed deviations from this mean-field solution.
    This procedure can be formally understood by replacing the operators within the Hamiltonian $\hat{H}$~\eqref{eq:supp:BH-Hamiltonian}
    via $\hat{a}_{\ell s} \to \sqrt{n_0} \zeta_{\ell s} + \hat{a}_{\ell s}$,
    where the first term denotes the complex-valued mean-field solution, $\zeta_{\ell s} \in \mathbb{C}$, and the second term
    the additional quantum fluctuations where $n_0$ denotes condensate 
    density of the system.
    Expanding the Bose-Hubbard Hamiltonian~\eqref{eq:supp:BH-Hamiltonian} in orders of the condensate 
    density $n_0$ then yields~\cite{Engelhardt2015} 
        \begin{align}
            \hat{H}= n_0 E_{\text{GP}}+n_0^{1/2} \hat{H}^{(L)}+\hat{H}^{(B)}+\mathcal{O}\left(n_0^{-1/2}\right).
            \label{eq:supp-General-Bog-Expansion-Dens}
        \end{align}
    The first term denotes the GP energy functional $ E_{\text{GP}}$ which is a function of the mean-field solutions
    $\zeta_{\ell s}$.
    The next two terms $\hat{H}^{(L)}$ and $\hat{H}^{(B)}$ are linear and quadratic in the bosonic operators
    $\hat{a}^{(\dagger)}_{\ell s}$, where the latter is known as the Bogoliubov Hamiltonian, which is sought after in the following.
    
    \subsection{Gross-Pitaevskii energy functional and equations}
    \label{subsec:GP-functional-TwoBand}
    First, the energy functional $E_{\text{GP}}$ is found by replacing the operators $\hat{a}^{(\dagger)}_{\ell s}$ in $\hat{H}$~\eqref{eq:supp:BH-Hamiltonian} by complex-valued functions $\zeta_{\ell s}^{(*)}$.
    We consider the case, where the condensate is formed at the single quasimomentum mode $\kv_0$ in the lower band, such that a suitable ansatz for the ground state is given
    by 
    \begin{align}
        \zeta_{\ell s} =\frac{e^{-i\kv_0 \cdot r_{\ell s}}}{\sqrt{N_{\text{uc}}}}\xi_s ,
        \label{eq:supp:MF-ansatz}
    \end{align}
    for $s \in \{A,B\}$.
    This ansatz represents a plane-wave condensate at quasimomentum $\kv_0$ with a sublattice structure given by the complex-valued functions $\xi_s$.
    If the condensate is formed at $\kv_0=\bm{0}$, the ansatz~\eqref{eq:supp:MF-ansatz} reduces to a spatially homogeneous plane-wave.
  
    This leads to the following energy functional in quasimomentum space
    \begin{align}
        E_{GP}=&\colvec{\xi_A^* \ \xi^*_B}(\mathcal{H}(\kv=\kv_0)-\mu_{\text{eff}}\mathbbm{1}_2)
        \colvec{\xi_A \\ \xi_B} \nonumber \\
        &+\frac{U}{2N_{\text{uc}}}\left(\abs{\xi_A}^4+\abs{\xi_B}^4 \right),
        \label{eq:supp:EGP-Functional}
    \end{align}
    where we have introduced $\mu_{\text{eff}}$ as a Lagrange multiplier for satisfying the 
    conservation of the total number of bosons in the system, i.e., $\sum_{\ell s} \abs{\zeta_{\ell s}}^2 = N$.
    We minimize $E_{GP}$ by taking the variational derivative of $E_{GP}$ with respect to $\xi_{s}^{*}$ which leads to 
    the so-called GP equations, 
        \begin{align}
            \big(\mathcal{H}(\kv=\kv_0)-\mu_{\text{eff}}\mathbbm{1}_2\big)\colvec{\xi_A\\ \xi_B} + \frac{U}{N_{\text{uc}}} \colvec{\abs{\xi_A}^2\xi_A\\ \abs{\xi_B}^2\xi_B}=0.
            \label{eq:supp:GPE-equations}
        \end{align}
    Since the non-interacting condensate is formed at the $\ket{\kv=\kv_0,-}$ mode we parametrize $\xi_s$ via the Bloch 
    sphere representation $\ket{\kv -}$~\eqref{eq:supp:kvec-plusmin-param}~\cite{Furukawa2015}
        \begin{align}
            \colvec{\xi_A\\ \xi_B} =\sqrt{N}\colvec{f_A \\f_B}= \sqrt{N}\colvec{\sin(\theta/2)\\ -\cos(\theta/2)e^{i\varphi}}.
            \label{eq:supp:Mean-field-param}
        \end{align}
    Note that this choice correctly adheres to the normalization condition $\sum_{\ell s} \abs{\zeta_{\ell s}}^2 = \sum_s \abs{\xi_s}^2 = N$.
    The GP equations in~\eqref{eq:supp:GPE-equations} give us four real equations but the unknowns we want to solve for are three real parameters, i.e., $\theta$, $\varphi$ and $\mu_{\text{eff}}$. 
    To solve this, we project the GP equations onto the two orthogonal vectors $\colvec{\xi_A^*\,\xi_B^*}$ and $\colvec{-\xi_B^*\,\xi_A^*}$, by multiplying~\eqref{eq:supp:GPE-equations} from the left. 
    These two equations then read
    \begin{align}
       & \rowvec{f_A^* & f_B^*}\big(\mathcal{H}(\kv=\kv_0)-\mu_{\text{eff}}\mathbbm{1}_2\big)\colvec{f_A\\ f_B} \nonumber \\
        &\quad + \frac{U}{N_{\text{uc}}}N \left(\abs{f_A}^4+\abs{f_B}^4 \right)=0,
        \label{eq:supp:GPE-proj1}\\
        &\rowvec{-f_B^* & f_A^*}\big(\mathcal{H}(\kv=\kv_0)-\mu_{\text{eff}}\mathbbm{1}_2\big)\colvec{f_A\\ f_B} \nonumber \\
        &\quad + \frac{U}{N_{\text{uc}}}N (-\abs{f_A}^2 f_A f_B^* +\abs{f_B}^2 f_A^* f_B)=0.
        \label{eq:supp:GPE-proj2}
    \end{align}
    Using the spherical parametrization of $H(\kv_0)$ as discussed in~\eqref{eq:supp:Hk-Pauli-exp}-\eqref{eq:supp:kvec-plusmin-param}
        \begin{align}
            H(\kv_0)&=h_0(\kv_0)\mathbbm{1}_2 + h(\kv_0)\,\hat{\bm{h}}(\kv_0) \cdot \bm{\hat{\sigma}} \nonumber \\
            &=\begin{pmatrix} h_0(\kv_0)+h(\kv_0)\cos\theta_0 & h(\kv_0)\sin\theta_0\,e^{-i\varphi_0} \\
            h(\kv_0)\sin\theta_0\,e^{i\varphi_0} & h_0(\kv_0)-h(\kv_0)\cos\theta_0 \end{pmatrix},
        \end{align}
    where $\theta_{0} \equiv \theta(\kv_0)$ and $\varphi_{0} \equiv \varphi(\kv_0)$ are the spherical angles of the non-interacting Hamiltonian at $\kv_0$.
    We now insert these expressions into~\eqref{eq:supp:GPE-proj1} and~\eqref{eq:supp:GPE-proj2}, which will, for the first, will yield a purely real and, for the second, a purely complex equation. The second equation can then be separated into two real equations by collecting real and imaginary parts, which have to be satisfied independently.
    Using straightforward algebra and a few trigonometric identities, we then arrive at the following set of three real equations for the three unknowns $\theta$, $\varphi$ and $\mu_{\text{eff}}$
    \begin{align}
        &(h_0(\kv_0)-\mu_{\text{eff}})-h(\kv_0)
        \big[\cos\theta_0\cos\theta \nonumber \\
        &\quad +\sin\theta_0\sin\theta\cos(\varphi-\varphi_0)\big]
        +2Un (\abs{f_A}^4 +\abs{f_B}^4)=0,
        \label{eq:supp:GPE-Min-eq1}\\
        &h(\kv_0)[\cos\theta_0\sin\theta\cos\varphi-\cos\varphi_0\sin\theta_0\cos\theta] \nonumber \\
        &\quad -Un\sin\theta\cos\theta\cos\varphi=0,
        \label{eq:supp:GPE-Min-eq2}\\
        &-(h_0(\kv_0)-\mu_{\text{eff}})\sin\theta\sin\varphi+h(\kv_0)\sin\theta_0\sin\varphi_0 \nonumber \\
        &\quad -Un\sin\theta\sin\varphi=0.
        \label{eq:supp:GPE-Min-eq3}
    \end{align}
    Here, we have defined the density $n=N/(2N_{\text{uc}})$, which is the number of bosons per lattice site~\cite{Furukawa2015,Wu2017}.
    For a given interaction energy $U$ and density $n$, we numerically minimize the
    \crefrange{eq:supp:GPE-Min-eq1}{eq:supp:GPE-Min-eq3} and gain the stationary solutions $(\theta,\varphi,\mu_{\text{eff}})$ specifying the mean-field ground state in~\eqref{eq:supp:Mean-field-param}.

    Since the BEC occurs in the lowest band at $\kv=\kv_0$, we further specify the unitary
    rotation (provided by the eigenstates of the non-interacting system~\eqref{eq:supp:kvec-plusmin-param})
    for the obtained solution~\cite{Furukawa2015}
            \begin{align}
                &\colvec{\hat{a}_{\kv_0, A}\\ \hat{a}_{\kv_0 , B}}
                = U(\theta,\varphi)
                \colvec{\hat{a}_{\kv_0 , +}\\ \hat{a}_{\kv_0 , -}}
                = \begin{pmatrix} f_{A,+} & f_{A,-} \\
                f_{B,+} & f_{B,-} \end{pmatrix}
            \colvec{\hat{a}_{\kv_0 , +}\\ \hat{a}_{\kv_0 , -}},
            \\[.2cm]
            &\text{with}\
            U(\theta,\varphi)
            = \begin{pmatrix} \cos(\theta/2) & \sin(\theta/2) \\
                \sin(\theta/2)e^{i\varphi} & -\cos(\theta/2)e^{i\varphi} \end{pmatrix},
            \end{align}
    where $f_{s,\pm}(\theta,\varphi)$ with $s \in \{A,B\}$ denote the stationary solutions 
    of the GP functional~\eqref{eq:supp:EGP-Functional}.

    \subsection{Bogoliubov Hamiltonian around the mean-field solution}
    \label{subsec:Bogoliubov-Hamiltonian-TwoBand}
    The next step is to go beyond this mean-field description and derive the corresponding Bogoliubov Hamiltonian
    $\hat{H}^{(B)}$~\eqref{eq:supp-General-Bog-Expansion-Dens}.
    The starting point is the quasimomentum space representation of the Bose-Hubbard Hamiltonian~\eqref{eq:supp:BH-Hamiltonian}.
    Utilizing another rotation into the eigenbasis~\eqref{eq:supp:kvec-plusmin-param}
    the bosonic operators in quasimomentum space $\hat{a}_{\kv , s} $ can be decomposed into a condensed and a non-condensed part via~\cite{Furukawa2015}
        \begin{align}
            \sum_{\kv}\hat{a}_{\kv, s} &= \hat{a}_{\kv_0, s} +\sum_{\kv \neq \kv_0}\hat{a}_{\kv, s} \nonumber \\
                &= f_{s,-}\hat{a}_{\kv_0, -}+f_{s,+}\hat{a}_{\kv_0, +}+\sum_{\kv \neq \kv_0}\hat{a}_{\kv, s},
                \label{eq:supp:Op-Cond-decomp}
            \end{align}
    where the condensation mode is assumed to be located at $\kv=\kv_0$ in the lowest band and
    $f_{s,\pm} \equiv f_{s,\pm}(\varphi,\theta)$ are functions depending on the stationary solutions of the energy functional
    $E_{GP}$~\eqref{eq:supp:EGP-Functional}.

    Next, the decomposition~\eqref{eq:supp:Op-Cond-decomp} is inserted into $\hat{H}$~\eqref{eq:supp:Ham-Full-2band-comb-momentum}
    and the Hamiltonian is separated into parts of $\kv=\kv_0$ and $\kv\neq\kv_0$ while only terms up to second order
    in $\hat{a}^{\dagger}_{\kv \neq \kv_0, s}$ are retained, leading to an approximate Hamiltonian.
    For the actual Bogoliubov approximation, the condensation mode is then replaced via $\hat{a}_{\kv_0, -} \to \sqrt{N_0}$,
    where $N_0$ denotes the number of particles in the condensate mode.
    Performing all these steps will eventually lead to the following Bogoliubov Hamiltonian in quasimomentum space
    \begin{widetext}
    \begin{align}
        \begin{split}
                \hat{H}^{(B)}=&
                \frac{1}{2}\sum_{\kv \neq \kv_0} 
                \rowvec{\hat{a}^\dg_{\kv_0+\kv,A} \, \hat{a}^\dg_{\kv_0+\kv,B}\,\hat{a}_{\kv_0-\kv,A},\hat{a}_{\kv_0-\kv,B}}
                \begin{pmatrix} \mathcal{H}(\kv_0+\kv)+\mathcal{H}_1 & 2\mathcal{H}_2 \\
                    2\mathcal{H}_{2}^{*}& (\mathcal{H}(\kv_0-\kv)+\mathcal{H}_{1})^{*}
                    \end{pmatrix}
                \colvec{\hat{a}_{\kv_0+\kv,A} \\ 
                \hat{a}_{\kv_0+\kv,B} \\
                \hat{a}^\dg_{\kv_0-\kv,A} \\ 
                \hat{a}^\dg_{\kv_0-\kv,B}}
                \\
                +&\frac{1}{2} 
                \rowvec{\hat{a}^\dg_{\kv_0,+} \, \hat{a}_{\kv_0,+}}
                \begin{pmatrix} h_{3}  & 2h_{4}  \\
                    2h_{4}^{*} & h_{3}^*
                \end{pmatrix}
                \colvec{\hat{a}_{\kv_0,+} \\ \hat{a}^\dg_{\kv_0,+}}
                +\text{const.}.
            \end{split}
                \label{eq:supp:Bogoliubov-Hamiltonian}
            \end{align}
    We can also compactly write this Bogoliubov Hamiltonian~\eqref{eq:supp:Bogoliubov-Hamiltonian}
    in terms of the Nambu spinor representation
            \begin{align}
            \hat{H}^{(B)}= \frac{1}{2}\sum_{\kv \neq \kv_0} \Psi_{\kv}^{\dagger}
                \mathcal{M}_{\kv} \Psi_{\kv}
                +\frac{1}{2}
                \Psi^{\dagger}_{+} 
                \mathcal{M}_{+}
                \Psi_{+}
                +\text{const.}, 
                \label{eq:supp:Bogoliubov-Hamiltonian-SpinorRep}
        \end{align}
    where $\Psi^{\dagger}_{\kv} \equiv \colvec{\hat{a}^{\dagger}_{\kv_0+\kv,A} \,\hat{a}^{\dagger}_{\kv_0+\kv,B}\,\hat{a}_{\kv_0-\kv,A},\hat{a}_{\kv_0-\kv,B}}$
    and $\Psi^{\dagger}_{+} \equiv \colvec{\hat{a}^{\dagger}_{\kv_0,+} \, \hat{a}_{\kv_0,+}}$.
    Here, $\mathcal{M}_+$ is the $2\times 2$ matrix and $\mathcal{M}_{\kv}$ 
    the deviation-momentum dependent $4\times 4$ coefficient matrix from~\eqref{eq:supp:Bogoliubov-Hamiltonian}, 
    which are explicitly given by~\cite{Furukawa2015,Wu2017}
    \begin{align}
        \begin{split}
                \mathcal{M}_{\kv} &= \begin{pmatrix} \mathcal{H}(\kv_0+\kv)+\mathcal{H}_1 & 2\mathcal{H}_2 \\
                2\mathcal{H}_{2}^{*}& (\mathcal{H}(\kv_0-\kv)+\mathcal{H}_{1})^{*}
                \end{pmatrix}, \quad
                \mathcal{M}_{+}= \begin{pmatrix} h_{3}  & 2h_{4}  \\
                    2h_{4}^{*} & h_{3}^*
                \end{pmatrix},\\[.3cm]
                \text{with} \
                \mathcal{H}_1&=4 Un \abs{F_{-}}^2
                -\mu_{\text{eff}} \mathbbm{1}_2,\quad
                \mathcal{H}_2=Un (F_{-})^2,\quad
            \end{split}
            \label{eq:supp:block-matrix}
        \end{align}
    where $F_{-}=\diag{f_{A,-},f_{B,-}}$ and 
    where the entries of $\mathcal{H}(\kv)$ are given by the explicit sublattice basis Hamiltonian~\eqref{eq:supp:Hk-Pauli-exp} which is determined by the underlying single-particle model.
    Note that for a condensate at $\kv_0=\bm{0}$ one has $\kv_0-\kv=-\kv$, so that~\eqref{eq:supp:block-matrix} reduces to the standard result of~\cite{Furukawa2015,Wu2017}. 
    For a finite condensation momentum $\kv_0\neq\bm{0}$ the lower-right block is evaluated at the partner momentum $\kv_0-\kv$, i.e., the reflection of $\kv_0+\kv$ about $\kv_0$. 
    Replacing $\kv_0-\kv$ by $-\kv$, which is the correct replacement only for $\kv_0=\bm{0}$, would yield an incorrect matrix that breaks the Goldstone zero mode.
    The entries of $M_{+}$ are given by 
            \begin{align}
                h_3=4Un \sum_{s}\left( \abs{f_{s,-}}^2\abs{f_{s,+}}^2\right) 
                 +\sum_{s,s'} f_{s,+}^{*}\mathcal{H}_{s,s'}(\kv_0)f_{s',+}-\mu_{\text{eff}}, 
                 \quad 
                h_4 = Un \sum_{s}(f_{s,-})^2 \,(f_{s,+}^*)^2,
            \end{align}
    where $\mu_{\text{eff}}$ follows the stationary solutions of \crefrange{eq:supp:GPE-Min-eq1}{eq:supp:GPE-Min-eq3}.
    The Bogoliubov Hamiltonian~\eqref{eq:supp:Bogoliubov-Hamiltonian} represents the symplectic Hopf model, which is the bosonic BdG generalization of the MRW Hopf model~\eqref{app:eq:hermitian-hopf-Hamiltonian}-\eqref{app:eq:hermitian-hopf-d-vectors}.

    The Bogoliubov Hamiltonian~\eqref{eq:supp:Bogoliubov-Hamiltonian} can now be diagonalized by applying 
    the paraunitary (Bogoliubov transformation) 
    $W_{\kv} \,(W_+)$~\cite{Furukawa2015} 
        \begin{align}
            \colvec{\hat{a}_{\kv_0+\kv,A} \\
                \hat{a}_{\kv_0+\kv,B} \\
                \hat{a}^\dg_{\kv_0-\kv,A} \\
                \hat{a}^\dg_{\kv_0-\kv,B}}
                = W_{\kv}
                \colvec{\hat{b}_{\kv_0+\kv,+} \\ \hat{b}_{\kv_0+\kv,-} \\
                \hat{b}^\dg_{\kv_0-\kv,+} \\ \hat{b}^\dg_{\kv_0-\kv,-}}
                \ \Leftrightarrow \ \Psi_{\kv}=W_{\kv} \beta_{\kv},
                \quad
                \colvec{\hat{a}_{\kv_0,+} \\ \hat{a}^\dg_{\kv_0,+}}
                =W_{+}
                \colvec{\hat{b}_{\kv_0,+} \\ \hat{b}^{\dagger}_{\kv_0,+}}
                \ \Leftrightarrow \
                \Psi_{+}=W_{+} \beta_{+}
            \label{eq:supp:Bog-Transf-simple}
        \end{align}
    with 
    $\beta_{\kv}\equiv \colvec{\hat{b}_{\kv_0+\kv,+} \, \hat{b}_{\kv_0+\kv,-} \,\hat{b}^\dg_{\kv_0-\kv,+} \, \hat{b}^\dg_{\kv_0-\kv,-}}^T$ 
    and $\beta_{+}\equiv \colvec{\hat{b}_{\kv_0,+} \, \hat{b}^{\dagger}_{\kv_0,+}}^T$
    consisting of the new bosonic quasi-particle annihilation (creation) operators 
    $\hat{b}^{(\dg)}_{\kv_0\pm\kv,\pm}$ at quasimomentum $\kv_0\pm\kv$ in the upper ($+$) or lower Bogoliubov energy band ($-$).
    The $4\times 4$ ($2\times 2$) Bogoliubov transformation matrix $W_{\kv}$ ($W_+$) can be parametrized as
    \begin{align}
        W_+ &=
        \begin{pmatrix}
            u_{\kv_0} & v_{\kv_0}^*\\
            v_{\kv_0} & u_{\kv_0}^*
        \end{pmatrix}, \quad 
            W_{\kv} = 
            \begin{pmatrix}
                U_{\kv_0+\kv} & V_{\kv_0-\kv}^*\\
                V_{\kv_0+\kv} & U_{\kv_0-\kv}^*
            \end{pmatrix}
            \ \
             \text{with} \ \
            U_{\kv_0+\kv} = \begin{pmatrix}
                u_{A,+}^{\kv_0+\kv} & u_{A,-}^{\kv_0+\kv}\\
                u_{B,+}^{\kv_0+\kv} & u_{B,-}^{\kv_0+\kv}
            \end{pmatrix}, 
            \ 
            V_{\kv_0+\kv} = \begin{pmatrix}
                v_{A,+}^{\kv_0+\kv} & v_{A,-}^{\kv_0+\kv}\\
                v_{B,+}^{\kv_0+\kv} & v_{B,-}^{\kv_0+\kv}
            \end{pmatrix}.
            \label{eq:supp:Bog-Transf-BogHald-paraunitary-param}
        \end{align}

    Since these quasi-particle operators still have to obey the bosonic commutation relations, 
    the Bogoliubov transformation matrix is paraunitary~\cite{Colpa1978,Shindou2013a,Tesfaye2025}, i.e., 
    $W^\dg \tau_z W=W\tau_z W^\dg =\tau_z,\  \text{with}\ \tau_z=\sigma_z \otimes \mathbbm{1}_2$ 
    ($\tau_z=\sigma_z$) for $W=W_{\kv}$ ($W=W_+$).
    These paraunitary matrices are specifically constructed such that
        \begin{align}
            W_{\kv}^{\dg}\mathcal{M}_{\kv} W_{\kv} = \diag{E_{+}(\kv_0+\kv),E_{-}(\kv_0+\kv),E_{+}(\kv_0-\kv),E_{-}(\kv_0-\kv)}, \quad
            W_{+}^{\dg}M_{+} W_{+} = \diag{E_{+}(\kv_0),E_{+}(\kv_0)},
        \end{align}
        holds,
        by which the Bogoliubov Hamiltonian is eventually diagonalized~\cite{Colpa1978,Shindou2013a,Furukawa2015,Tesfaye2025} to
        \begin{align}
            \hat{H}^{(B)}=&\sum_{\kv} E_{+}(\kv)\hat{b}^{\dagger}_{\kv,+} \hat{b}_{\kv,+}+\sum_{\kv \neq \kv_0} E_{-}(\kv)\hat{b}^{\dagger}_{\kv,-} \hat{b}_{\kv,-} +\text{const.}\,.
            \label{eq:supp:Bogoliubov-Hamiltonian-Diag}
        \end{align}
    Here, $E_{\pm}(\kv)$ is the Bogoliubov energy spectrum of the upper ($+$) and lower ($-$) band evaluated at quasimomentum $\kv$.
    
    Note that the correct Bogoliubov energy spectrum can be obtained by numerically solving the energy eigenvalue problem of the (non-hermitian) dynamical matrix $D(\kv)=\tau_z \mathcal{M}_{\kv}$, i.e., $D(\kv)\ket{w_n(\kv)} =\tilde{\Omega}_n \ket{w_n(\kv)}$, while the correct paraunitary Bogoliubov transformation matrix $W_{\kv}$ is \emph{not} obtained by the corresponding eigenvectors, but rather requires a specific construction procedure involving a Cholesky decomposition of the coefficient matrices $\mathcal{M}_{\kv}$ and $M_{+}$ as described in detail in~\cite{Colpa1978,Shindou2013a,Tesfaye2025}.
        
    \end{widetext}
\section{Review of bosonic Bogoliubov-de Gennes (BdG) Hamiltonians and indefinite inner product spaces}
\label{app:review-Bogoliubov-Hamiltonian-Indefinite-Inner-Product}
The Bogoliubov Hamiltonian~\eqref{eq:supp:Bogoliubov-Hamiltonian-SpinorRep} is quadratic in bosonic Nambu operators and its diagonalization must preserve the bosonic commutation relations. 
We summarize here the algebraic structure of the underlying indefinite (Krein) inner product space that the main text uses~(see~\Ccite{Colpa1978,Shindou2013a,Tesfaye2025,Furukawa2015} for more details).

\subsection{Indefinite inner product and Krein space}
\label{app:subsec:rev-krein-space}

The Nambu coefficient space $\mathbb{C}^4$ is equipped with the indefinite inner product $\bra{\psi}\tau_z\ket{\psi}$, with the Krein metric $\tau_z = \sigma_z\otimes\mathbbm{1}_2$, which is positive on the particle sector and negative on the hole sector. A vector is Krein-positive if $\bra{\psi}\tau_z\ket{\psi} > 0$ and Krein-negative if $\bra{\psi}\tau_z\ket{\psi} < 0$. The pair $(\mathbb{C}^4,\tau_z)$ is a Krein space of signature $(2,2)$ for two bosonic modes per unit cell, generalizing to signature $(n,n)$ for $n$ modes~\cite{Peano2018,Schulz-Baldes2017}.

The Bogoliubov transformation $\Psi = W\beta$ that diagonalizes $\hat{H}^{(B)}$ preserves the bosonic commutation relations $[\Psi_i,\Psi_j^\dg] = \tau_{ij}$ if and only if $W$ is paraunitary~\eqref{eq:paraunitary}, $W^\dg\tau_z W = W\tau_z W^\dg = \tau_z$. The fermionic Bogoliubov problem instead has a positive-definite Nambu metric and requires unitarity, $W^\dg W = \mathbbm{1}$. The indefinite $\tau_z$ is a direct consequence of bosonic statistics with no fermionic counterpart.

\subsection{Dynamical matrix and pseudo-Hermiticity}
\label{app:subsec:rev-dynamical}

Diagonalizing $\hat{H}^{(B)}$~\eqref{eq:supp:Bogoliubov-Hamiltonian-SpinorRep} reduces to the eigenvalue problem of the dynamical matrix $D(\kv) = \tau_z\,\mathcal{M}_{\kv}$~\eqref{eq:dyn-matrix-main}. Since $\mathcal{M}_{\kv} = \mathcal{M}_{\kv}^\dg$~\eqref{eq:supp:block-matrix} and $\tau_z^2 = \mathbbm{1}$, the matrix $D$ obeys the $\tau_z$-pseudo-Hermiticity $D^\dg = \tau_z D\tau_z$. Pseudo-Hermiticity constrains the spectrum but does not by itself make it real.

\emph{Stability (Colpa's theorem~\cite{Colpa1978}).} If $\mathcal{M}_{\kv} > 0$ (positive definite), then $D = \tau_z\mathcal{M}$ is similar to the Hermitian matrix $\mathcal{M}^{1/2}\tau_z\mathcal{M}^{1/2}$ of signature $(2,2)$, so $D$ is diagonalizable with a real spectrum of two positive and two negative eigenvalues. From the generalized eigenvalue equation $\mathcal{M}_{\kv}\ket{\psi} = E\tau_z\ket{\psi}$ together with $\bra{\psi}\mathcal{M}_{\kv}\ket{\psi} > 0$, one reads off $E\bra{\psi}\tau_z\ket{\psi} > 0$, so
\begin{align}
    \mathrm{sign}(E) = \mathrm{sign}\bigl(\bra{\psi}\tau_z\ket{\psi}\bigr),
    \label{app:eq:rev-krein-sign}
\end{align}
i.e., the particle (positive-energy) Bogoliubov bands are exactly the Krein-positive eigenvectors, a consequence of the stability condition. When $\mathcal{M}_{\kv} > 0$ fails, pairs of eigenvalues can collide and leave the real axis, a bosonic exceptional-point instability with no fermionic counterpart~\cite{Bergholtz2021}.

\subsection{Pseudo-Hermitian (Riesz) projectors}
\label{app:subsec:rev-riesz}

Because $D(\kv)$ is pseudo-Hermitian, smooth band eigenvectors across $T^3$ are constructed from the Riesz (contour) projector~\cite{Schulz-Baldes2017,Peano2018,Lein2019}
\begin{align}
    P(\kv) = \frac{1}{2\pi i}\oint_\Gamma \bigl(z - D(\kv)\bigr)^{-1}\,dz,
    \label{app:eq:rev-riesz}
\end{align}
where $\Gamma$ encloses the band of interest and no other eigenvalue. A uniformly open BdG gap on $T^3$ lets $\Gamma$ be chosen $\kv$-independent, so $P(\kv)$ is smooth. The pseudo-Hermiticity of $D$ carries over to the projector,
\begin{align}
    \tau_z P(\kv)^\dg\tau_z = P(\kv). 
    \label{app:eq:rev-riesz-krein}
\end{align}
It is the pseudo-Hermitian counterpart of the orthogonal spectral projector $\projector{u}$ of the Hermitian theory, and is the projector entering the spectral evaluation of the symplectic Berry curvature in~\cref{app:numerical-hopf-invariant}.

\section{Quantization of the symplectic Hopf invariant}
\label{app:quantization}

We establish that $\chi$ is integer-valued and a complete homotopy invariant under stability~(S) and gap~(G) conditions, and prove that the necessary vanishing condition~(W) holds automatically on the phase diagram of \cref{fig:phase-diagram-hopf-bdg}. 
We work at $\kv_0 = \bm{0}$ and $n=2$ bosonic modes per unit cell throughout.

\subsection{Classifying space and deformation retraction}
\label{app:subsec:quant-classifying}

\emph{Claim 1.} The Krein-positive projective domain
\begin{align}
    \mathcal{L}^+_n = \bigl\{[\psi]\in\mathbb{CP}^{2n-1} : \bra{\psi}\tau_z\ket{\psi} > 0\bigr\},
    \label{app:eq:quant-classifying}
\end{align}
is an open, non-compact subset of $\mathbb{CP}^{2n-1}$, homeomorphic to $U(n,n)/(U(1)\times U(n-1,n))$ as a homogeneous space. 
For $n=2$: $\mathcal{L}^+_2 \cong U(2,2)/(U(1)\times U(1,2))\simeq\mathbb{CP}^1 \cong S^2$. The domain $\mathcal{L}^+_n$ also strong-deformation-retracts onto $\mathbb{CP}^{n-1}$.

\emph{Proof (explicit retraction).} Write $\ket{\psi} = (u,v)^T$ with $u,v\in\mathbb{C}^n$. Krein-positivity $|u|^2 - |v|^2 > 0$ forces $u\neq 0$. Define the homotopy
\begin{align}
    r_t\bigl([(u,v)]\bigr) = \bigl[(u,\,(1-t)v)\bigr], \quad t\in[0,1].
    \label{app:eq:quant-retraction}
\end{align}
This is well defined on projective classes and continuous in $t$. 
It stays in $\mathcal{L}^+_n$: $|u|^2 - (1-t)^2|v|^2 \geq |u|^2 - |v|^2 > 0$. 
At $t=0$, $r_0 = \mathrm{id}$, at $t=1$, $r_1([(u,v)]) = [(u,0)]\in\mathbb{CP}^{n-1}\hookrightarrow\mathcal{L}^+_n$~\footnote{The hooked arrow $\hookrightarrow$ denotes an inclusion: $\mathbb{CP}^{n-1}$ sits inside $\mathcal{L}^+_n$ as a subspace, here the $v=0$ locus $\{[(u,0)]\}$.} and $r_t$ fixes $\mathbb{CP}^{n-1}$ pointwise for all $t$. 
This is therefore a strong deformation retraction.

\subsection{Existence of the Hopf invariant}
\label{app:subsec:quant-existence}

\emph{Claim 2.} $\pi_3(\mathcal{L}^+_n) = \pi_3(\mathbb{CP}^{n-1})$, which equals $\mathbb{Z}$ for $n=2$ and $0$ otherwise.

\emph{Proof.} For $n=1$: $\mathbb{CP}^0$ is a point, so $\pi_3 = 0$. 
For $n\geq 2$: the Hopf fibration $S^1\hookrightarrow S^{2n-1}\to\mathbb{CP}^{n-1}$ gives a long exact sequence in homotopy. 
The relevant segment is $\pi_3(S^1)\to\pi_3(S^{2n-1})\to\pi_3(\mathbb{CP}^{n-1})\to\pi_2(S^1)$, and since $\pi_3(S^1) = \pi_2(S^1) = 0$, one obtains $\pi_3(\mathbb{CP}^{n-1})\cong\pi_3(S^{2n-1})$. For $n=2$: $\pi_3(S^3) = \mathbb{Z}$, since $\pi_m(S^m) = \mathbb{Z}$. 
For $n\geq 3$: $2n-1\geq 5$, and $\pi_3(S^{2n-1}) = 0$ since $\pi_k(S^m) = 0$ for $k<m$~\cite{Nakahara2003,Kennedy2015}. 

A symplectic Hopf invariant therefore exists if and only if the BBdG system has exactly two bosonic modes (per unit cell), i.e., a $4\times4$ BdG matrix. 
This is the exact bosonic counterpart of the non-stable character of the fermionic Hopf insulator, which also requires exactly two bands. 
By contrast, $\pi_2(\mathbb{CP}^{n-1}) = \mathbb{Z}$ for all $n\geq 2$, so symplectic Chern numbers persist as weak invariants wherever the Hopf invariant does not exist.

Notice that the above applies to a single isolated Bogoliubov band. 
A rank-$r$ Krein-positive multiplet has a Krein Grassmannian as classifying space, retracting to $\mathrm{Gr}_r(\mathbb{C}^n)$. 
Any Hopf-type invariant in that generalized setting is beyond the scope of this work.

\subsection{The mode-number family}
\label{app:subsec:quant-modes}

The Bogoliubov construction yields a family of $U(1)$ bundles over Krein-positive projective domains, one for each mode number $n$. 
The total space of Krein-unit vectors is $E_n = \{\ket{\psi} : \bra{\psi}\tau_z\ket{\psi} = 1\}$. 
Given $v\in\mathbb{C}^n$ arbitrary, $u$ ranges over a sphere of radius $\sqrt{1+|v|^2}$, so $E_n\cong S^{2n-1}\times\mathbb{R}^{2n}$, which is non-compact. 
Dividing out the overall $U(1)$ phase returns the base space, $E_n/U(1) = \mathcal{L}^+_n$, the Bogoliubov counterpart of the usual Berry-phase $U(1)$ over the Brillouin zone.

\begin{table}[t]
\centering
\small
\begin{tabular}{ccccp{2cm}}
\toprule
modes $n$ & total space $E_n$ & base $\mathcal{L}^+_n$ & $\pi_3(\text{base})$ & identification \\
\midrule
$1$ & $S^1\times\mathbb{R}^2$ & $\mathbb{D}\simeq\mathrm{pt}$ & $0$ & non-compact first Hopf map~\cite{Hasebe2010,Hasebe2010a} \\
$2$ & $S^3\times\mathbb{R}^4\simeq S^3$ & $\simeq S^2$ & $\mathbb{Z}$ & symplectic Hopf map (this work) \\
$\geq 3$ & $\simeq S^{2n-1}$ & $\simeq\mathbb{CP}^{n-1}$ & $0$ & no Hopf invariant \\
\bottomrule
\end{tabular}
\caption{Family of $U(1)$ bundles over Krein-positive projective domains for $n$ bosonic modes per unit cell. 
The $n=1$ member is Hasebe's non-compact first Hopf map, $n=2$ is the symplectic Hopf map of this work while $n\geq3$ has trivial $\pi_3$.}
\label{tab:quant-modes}
\end{table}

The total space $E_2$ retracts to $S^3$, and quotienting by the $U(1)$ phase then yields the base $S^2$. 
Under the retraction $v\to 0$, the symplectic Hopf map becomes the ordinary Hopf fibration $S^3\to S^2$. 
That is the precise sense in which $\chi$ is a Hopf invariant. 
Hasebe's non-compact Hopf maps~\cite{Hasebe2010,Hasebe2010a} and the symplectic Hopf map thus belong to one family of $U(1)$ bundles over Krein-positive projective domains, indexed by the mode number, with non-trivial $\pi_3$ for exactly one member.
The group-theoretically precise name for $\mathcal{L}^+_n$ is a pseudo-unitary or Krein projective domain. 
The adjective ``symplectic'' here follows the paraunitary convention of the BBdG literature, as discussed in \cref{sec:symplectic-hopf-model}.

\subsection{The symplectic connection is a genuine $U(1)$ connection}
\label{app:subsec:quant-connection}
\emph{Claim 3.} $\mathcal{A}_\mu = i\bra{w_+}\tau_z\ket{\del_\mu w_+}$, with $\bra{w_+}\tau_z\ket{w_+} = +1$, is a real-valued $U(1)$ connection on $T^3$.

\emph{Proof.} 
\emph{Reality:} differentiating $\bra{w_+}\tau_z\ket{w_+} = 1$ with respect to $k_\mu$ gives $2\,\re{\bra{w_+}\tau_z\ket{\del_\mu w_+}} = 0$, so $\mathcal{A}_\mu\in\mathbb{R}$. 
This positive normalization can be fixed only because $\ket{w_+}$ is Krein-positive (condition~(S)); without it $\mathcal{A}$ would not be real. 
\emph{Gauge transformation:} $\ket{w_+}\to e^{i\theta(\kv)}\ket{w_+}$ leads to $\mathcal{A}\to\mathcal{A} - d\theta$.

Hence, $\mathcal{A}$ is the Chern connection of the tautological $U(1)$ line bundle over $\mathcal{L}^+_2$, pulled back by $f: T^3\to\mathcal{L}^+_2$. 
Condition~(S) fixes the Krein normalization $\bra{w_+}\tau_z\ket{w_+} = +1$ to a positive constant, and this is exactly what lets $\mathcal{A}$ and its curvature $\mathcal{F} = d\mathcal{A}$ be defined smoothly everywhere on $\mathcal{L}^+_2$, just as for the Berry connection of an ordinary Bloch band.
Consequently, $[\mathcal{F}/2\pi] = f^*c_1\in H^2(T^3;\mathbb{Z})\cong\mathbb{Z}^3$, giving the three weak symplectic Chern numbers $C_{yz}, C_{zx}, C_{xy}$.

\subsection{The weak Chern numbers vanish}
\label{app:subsec:quant-weak-chern}

\emph{Claim 4.} On every connected component of the stable gapped region that meets the $Un = 0$ axis, $C_{yz} = C_{zx} = C_{xy} = 0$.
The Whitehead integral requires condition~\textbf{(W)}: $\alpha \equiv f^*c_1 = 0$.

\emph{Proof.} 
(1) $\alpha = f^*c_1\in H^2(T^3;\mathbb{Z})\cong\mathbb{Z}^3$ is a homotopy invariant of $f$, valued in a discrete group. 
(2) The Riesz projector construction~\eqref{app:eq:rev-riesz} makes $f$ depend continuously on $(m,Un)$ throughout the stable gapped region: the contour $\Gamma$ can be held fixed as long as no eigenvalue crosses it, and a continuous map into a discrete group is locally constant. 
Hence, $\alpha$ is constant on each connected component of the stable gapped region. 
(3) At $Un = 0$ the off-diagonal block $\mathcal{H}_2 = Un F_-^2 = 0$ vanishes, $\mathcal{M}_{\kv}$ is block-diagonal, and the particle eigenvector reduces to $(z(\kv),0)^T$, where $z(\kv)$ is the MRW Bloch spinor~\eqref{eq:supp:kvec-plusmin-param}. 
The map $f$ factors as $T^3\xrightarrow{z}S^3\xrightarrow{\eta_1}\mathbb{CP}^1\cong S^2$, where $\eta_1$ is the standard Hopf fibration. 
Thus, $\alpha = z^*\eta_1^*c_1$, and $\eta_1^*c_1\in H^2(S^3;\mathbb{Z}) = 0$, so $\alpha = 0$. 
(4) By local constancy, $\alpha\equiv 0$ on every connected component of the stable gapped region containing a point on the $Un = 0$ axis. 
In~\cref{fig:phase-diagram-hopf-bdg} this covers the $\chi = -2$, $+1$ and $0$ regions.

\subsection{Quantization}
\label{app:subsec:quant-quantization}

\emph{Claim 5.} Under conditions~(S), (G), and~(W), the symplectic Hopf invariant
\begin{align}
    \chi = -\frac{1}{4\pi^2}\int_{T^3}\mathcal{A}\wedge\mathcal{F}
    \label{app:eq:quant-hopf-invariant}
\end{align}
is integer-valued and is a complete invariant of the homotopy class of $f: T^3\to\mathcal{L}^+_2$.

\emph{Well-definedness.} The Whitehead integral~\eqref{app:eq:quant-hopf-invariant} is gauge invariant, hence independent of the choice of potential $\mathcal{A}$, precisely when the weak Chern numbers vanish (condition~(W), Claim~4). 
Otherwise $\chi$ is defined only modulo $2\gcd(C_{yz},C_{zx},C_{xy})$, the Hopf-Chern situation of~\cite{Kennedy2016}. 

\emph{Integrality and completeness.} 
By Claim~1 (retraction to $S^2$) and smooth approximation, $f$ is homotopic to a smooth map $T^3\to S^2$. 
Pontryagin's classification of such maps~\cite{Pontryagin1941} labels them by their weak Chern data $\alpha$ and, at fixed $\alpha$, by one additional (secondary) integer invariant.
With $\alpha = 0$ (Claim~4) this secondary invariant is unambiguous and takes any value in $H^3(T^3;\mathbb{Z}) = \mathbb{Z}$. 
The Whitehead integral~\cite{Whitehead1947}~\eqref{app:eq:quant-hopf-invariant} computes exactly this secondary invariant, giving integrality and completeness simultaneously.

\subsection{The Goldstone point}
\label{app:subsec:quant-goldstone}
In the main text we consider only the upper particle Bogoliubov band. Here we briefly discuss the lower band, where a Goldstone mode appears at $\kv_0$, and its relation to the symplectic Hopf invariant. 
The Bogoliubov spectrum has a gapless Goldstone mode at $\kv = \kv_0 = \bm{0}$, where $E_-(\kv_0) = 0$. 
For $Un > 0$, both particle and hole components of the Bogoliubov mode diverge as $|\kv|^{-1/2}$ near $\kv_0$, while the Krein norm $|u(\kv)|^2 - |v(\kv)|^2 = 1$ stays fixed. 
Consequently, $|u(\kv)|/|v(\kv)|\to 1$ and the projective point $[(u,v)]$ approaches the Krein-null boundary $\partial\mathcal{L}^+_2 = \{|u|^2 = |v|^2\}$ as $\kv\to\kv_0$. 
The map $f: T^3\to\mathcal{L}^+_2$ does not extend continuously to $\kv_0$, and $\mathcal{A}$ is singular there.

\emph{Claim 6.} 
The retracted texture removes this singularity. 
As $\kv\to\kv_0$, the fluctuation is along the condensate direction, so $u(\kv)$ approaches the condensate spinor $\xi = (f_A,f_B)^T$~\eqref{eq:supp:Mean-field-param} up to normalization. 
So one can define an effective Hopf texture
\begin{align}
    \bm{d}_{\mathrm{eff}}(\kv_0) = \frac{u(\kv_0)^\dg\,\bm{\hat\sigma}\,u(\kv_0)}{|u(\kv_0)|^2} = \frac{\xi^\dg\,\bm{\hat\sigma}\,\xi}{|\xi|^2}
    \label{app:eq:quant-goldstone-limit}
\end{align}
which is the condensate Bloch vector, a regular point of $S^2$. 
The block $u(\kv)$ is nonzero everywhere on $T^3$: for $\kv\neq\kv_0$ this follows from Krein positivity $|u|^2 > |v|^2\geq 0$, and at $\kv_0$ from the Goldstone-mode limit just described. 
Therefore, $\bm{d}_{\mathrm{eff}}: T^3\to S^2$ is continuous on all of $T^3$, including $\kv_0$. 

The symplectic Hopf invariant thus equals
\begin{align}
    \chi = \mathrm{Hopf}(\bm{d}_{\mathrm{eff}}),
    \label{app:eq:quant-deff-hopf}
\end{align}
where $\mathrm{Hopf}(\bm{d}_{\mathrm{eff}})$ is the ordinary Hopf invariant of $\bm{d}_{\mathrm{eff}}: T^3\to S^2$, computed by the Whitehead formula~\eqref{app:eq:hopf-invariant}.
The equality follows from homotopy invariance: $f\simeq\bm{d}_{\mathrm{eff}}$ as maps $T^3\to S^2$ after the retraction, so the two integers agree even though the two integrands differ pointwise (the symplectic connection $\mathcal{A}$ carries the $|v|\neq 0$ dressing from the Bogoliubov transformation). 
The $\bm{d}_{\mathrm{eff}}$ formulation is therefore what makes $\chi$ well-defined even at the Goldstone point.

\subsection{Particle-hole symmetry of the invariant}
\label{app:subsec:quant-ph}
\emph{Claim 7.} $\chi^{(\mathrm{hole})} = -\chi^{(\mathrm{particle})}$.

\emph{Proof.} 
We verify the particle-hole structure directly from~\eqref{eq:supp:block-matrix}. With $\tau_x = \sigma_x\otimes\mathbbm{1}_2 = \bigl(\begin{smallmatrix}0&\mathbbm{1}_2\\\mathbbm{1}_2&0\end{smallmatrix}\bigr)$, write $\mathcal{M}_{\kv}$ in block form with $A(\kv) = \mathcal{H}(\kv)+\mathcal{H}_1$ and $B = 2\mathcal{H}_2$:
\begin{align}
    \mathcal{M}_{\kv} = \begin{pmatrix} A(\kv) & B \\ B^* & A(-\kv)^* \end{pmatrix}.
    \label{app:eq:quant-M-block}
\end{align}
A direct block computation gives $\mathcal{M}_{-\kv}^* = \bigl(\begin{smallmatrix}A(-\kv)^* & B^*\\B & A(\kv)\end{smallmatrix}\bigr)$, and therefore
\begin{align}
    \tau_x\,\mathcal{M}_{-\kv}^*\,\tau_x
    = \begin{pmatrix} A(\kv) & B \\ B^* & A(-\kv)^* \end{pmatrix}
    = \mathcal{M}_{\kv},
    \label{app:eq:quant-ph-Msymmetry}
\end{align}
holding for any single-particle Hamiltonian $\mathcal{H}(\kv)$, provided the pairing block $B$ is momentum-independent, as is the case here since the on-site interaction $\mathcal{H}_2$ carries no $\kv$-dependence (a $\kv$-dependent pairing block would additionally require $B(-\kv)=B(\kv)$). Since $\tau_x\tau_z\tau_x = -\tau_z$,
\begin{align}
    D(\kv) = \tau_z\mathcal{M}_{\kv} = -\tau_x\,D(-\kv)^*\,\tau_x.
    \label{app:eq:quant-ph-D}
\end{align}

If $D(-\kv)\ket{w_p(-\kv)} = E\ket{w_p(-\kv)}$ with $E > 0$ (particle band, Krein sign $s_p = +1$), then $D(\kv)\ket{w_h(\kv)} = -E\ket{w_h(\kv)}$ with
$\ket{w_h(\kv)} = \tau_x(\ket{w_p(-\kv)})^* = \bigl(v^*(-\kv),\,u^*(-\kv)\bigr)^T$,
where $\ket{w_p} = (u,v)^T$ splits into the particle block $u$ and the hole block $v$.
Its Krein norm $\bra{w_h(\kv)}\tau_z\ket{w_h(\kv)} = |v^*(-\kv)|^2 - |u^*(-\kv)|^2 = -\bigl(|u(-\kv)|^2 - |v(-\kv)|^2\bigr) = -1$ is negative, so $w_h$ has the hole Krein sign $s_h = -1$, as it must once condition~(S) fixes the particle norm $\bra{w_p}\tau_z\ket{w_p} = +1$. 
Retracting the hole vector (killing the upper block in $\mathcal{L}^-$) gives $\bm{d}^{(\mathrm{hole})}_{\mathrm{eff}}(\kv) = (\bm{d}^{(\mathrm{particle})}_{\mathrm{eff}}(-\kv))^*$, i.e., BZ inversion on the base composed with complex conjugation on the target $S^2$.

Three independent factors determine the sign of $\chi^{(\mathrm{hole})}$ relative to $\chi^{(\mathrm{particle})}$:
\begin{enumerate}
    \item \emph{Base inversion.} 
    The reparametrization $\kv\to-\kv$ of $T^d$ has orientation sign $(-1)^d$, equal to $-1$ in $d=3$ (Hopf) and $+1$ in $d=2$ (Chern).
    \item \emph{Target conjugation.} 
    Complex conjugation on $S^2$ is an orientation-reversing reflection of degree $-1$. 
    An invariant linear in the connection (the Chern number $\propto\!\int\mathcal{F} = \int d\mathcal{A}$) inherits this degree once, giving $-1$.
    Here, the Hopf invariant is quadratic in the connection (the abelian Chern-Simons form $\propto\!\int\mathcal{A}\wedge\mathcal{F} = \int\mathcal{A}\wedge d\mathcal{A}$, equivalently scaling as the square of the target degree) giving $(-1)^2 = +1$.
    \item \emph{Hole Krein normalization.} 
    The sign $s_h = -1$ enters through the normalized connection $\mathcal{A}_\mu = i\bra{w}\tau_z\ket{\del_\mu w}/\bra{w}\tau_z\ket{w}$ in the same linear-versus-quadratic way: once for the Chern number, giving $-1$, and squared for the Hopf invariant, giving $+1$.
\end{enumerate}
For the symplectic Hopf invariant ($d=3$) the three factors multiply to $(-1)(+1)(+1) = -1$, which establishes $\chi^{(\mathrm{hole})} = -\chi^{(\mathrm{particle})}$ and proving Claim~7.

The contrast with the $d=2$ symplectic Chern result $C^{\mathrm{symp}}_{\mathrm{hole}} = +C^{\mathrm{symp}}_{\mathrm{particle}}$~\cite{Shindou2013a,Furukawa2015} becomes clear. 
The same three factors now read $(+1)(-1)(-1) = +1$, since the inversion $\kv\to-\kv$ is orientation-preserving in $d=2$ while the target conjugation and the Krein sign each enter the linear Chern integrand only once. 
The two Chern numbers are therefore equal, consistent with the paraunitary Chern sum rule established for thermodynamically stable BBdG systems in~\cite{Shindou2013a}. 

\section{Open boundary symplectic Hopf edge states}
\label{app:edge-states}
We provide some details regarding the open-boundary computation of \cref{subsec:edge-states} in the main text.

To expose surface states we open the cubic lattice along $z$ while keeping $\kv_\parallel = (k_x, k_y)$ periodic. Writing the bulk BBdG coefficient matrix as a Fourier series in the opened Bloch phase $k_z$,
\begin{align}
    \mathcal{M}(\kv_\parallel, k_z) = \sum_{n=-n_{\max}}^{n_{\max}} t_n(\kv_\parallel)\,e^{-i k_z n},
    \label{app:eq:edge-hopping}
\end{align}
the inter-layer coefficient matrices $t_n$ are recovered exactly by an inverse discrete Fourier transform, $t_n = \mathrm{IDFT}_{k_z}[\mathcal{M}]_n$, provided the sampling meets the Nyquist bound $N_{k_z} \geq 2 n_{\max} + 1$, which is essentially the method of Ref.~\cite{Deng2013} adapted to the para-unitary setting.
For the present model the $z$-hoppings are nearest-neighbour only ($n_{\max} = 1$), so any $N_{k_z} \geq 3$ is exact and independent of the real-space slab thickness $L_z$. 
Stacking $L_z$ layers with open boundaries gives the block-banded slab coefficient matrix $[\mathcal{M}_{\mathrm{slab}}]_{ab} = t_{b-a}$, $a, b = 0, \dots, L_z - 1$. 
It is diagonalized through its dynamical matrix $D_{\mathrm{slab}} = \tau_z \mathcal{M}_{\mathrm{slab}}$ exactly as in the bulk (\cref{app:review-Bogoliubov-Hamiltonian-Indefinite-Inner-Product}), yielding signed energies $E_j$ and modes $\ket{w_j} = (u_j, v_j)^T$, where $j = 1, \dots, 2 L_z$ labels the slab modes.

\bibliographystyle{apsrev4-2-arxiv-notitlelink}
\bibliography{main}  %
\end{document}